\documentclass{new_tlp}
\usepackage{xspace}
\usepackage[utf8]{inputenc}
\usepackage[T1]{fontenc}
\usepackage{proof}
\usepackage{xcolor}
\usepackage{amsmath}
\usepackage{amssymb}
\usepackage{cmll} 
\definecolor{darkblue}{RGB}{0, 0, 185}

\usepackage{hyperref} 
\hypersetup{
     colorlinks=true,
     linkcolor=darkblue,
     filecolor=darkblue,
     citecolor = darkblue,
     urlcolor=darkblue,
 }
\usepackage{breakurl}

\newcommand{\tup}[1]{\langle #1\rangle}

\newcommand{\lP}{$\lambda$Prolog\xspace}
\newcommand{\Dscr}{{\mathcal D}}
\newcommand{\Fscr}{{\mathcal F}}
\newcommand{\Gscr}{{\mathcal G}}

\newcommand{\Pscr}{{\mathcal P}}
\newcommand{\eg}{e.g.}
\newcommand{\LK }{\ensuremath{\mathbf{LK }}\xspace}
\newcommand{\LKF}{\ensuremath{\mathbf{LKF}}\xspace}
\newcommand{\LL }{\ensuremath{\mathbf{LL }}\xspace}
\newcommand{\LLp}{\ensuremath{\mathbf{LL}'}\xspace}
\newcommand{\LLf}{\ensuremath{\mathbf{LL^f}}\xspace}
\newcommand{\LJ }{\ensuremath{\mathbf{LJ }}\xspace}
\newcommand{\LJF}{\ensuremath{\mathbf{LJF}}\xspace}
\newcommand{\LJFp}{\ensuremath{\mathbf{LJF}'}\xspace}

\newcommand{\Seq}[2]{#1\vdash #2}
\newcommand{\Seqq}[3]{#1\colon\,#2\vdash #3}
\newcommand{\lSeq}[3]{\Seq{#1\;;#2}{#3}}
\newcommand{\lSeqq}[4]{\Seqq{#1}{#2\;;#3}{#4}}

\newcommand{\fSeqq}[5]{\Seq{#1\colon\,#2\;;#3\mathrel{\DOWNarrow}#4}{#5}}

\newcommand{\adj}[2]{\hbox{\sl adj}~#1~#2}
\newcommand{\pth}[2]{\hbox{\sl path}~#1~#2}

\newcommand{\DOWNarrow}{{\color{darkblue}\Downarrow}}
\newcommand{\jLf }[3]{#1\mathbin\DOWNarrow\null#2\vdash#3} 
\newcommand{\jRf }[2]{#1\vdash #2\mathbin\DOWNarrow\null}  
\newcommand{\labelleft}[1]{\deduce[\hbox{#1}]{\strut}{}}
\newcommand{\typeof}[2]{\hbox{\sl typeof}~#1~#2}
\newcommand{\ra}{\rightarrow}

\newcommand{\impR}{\imp R}
\newcommand{\impL}{\imp L}
\newcommand{\forallR}{\forall R}
\newcommand{\forallL}{\forall L}

\newcommand{\contrL}{contr\xspace}
\newcommand{\weakL}{weak\xspace}
\newcommand{\init}{init\xspace}
\newcommand{\cut}{cut\xspace}
\newcommand{\absorb}{absorb\xspace}
\newcommand{\decide}{\hbox{decide}\xspace}
\newcommand{\decideb}{\hbox{decide}\bang\xspace}

\newcommand{\backchain}{\hbox{backward chaining}\xspace}
\newcommand{\dereliction}{\hbox{dereliction}\xspace}
\newcommand{\limpL}{\limp L}
\newcommand{\limpR}{\limp R}

\newcommand{\bang}{!}

\newcommand{\limp}{\multimap}

\newcommand{\imp}{\supset}
\newcommand{\sep}{\mathrel{|}}
\newcommand{\truth}{\top}
\newcommand{\false}{\bot}

\def\Strut{\vrule height.7\baselineskip depth0.5\baselineskip width0.0pt}

\newcommand{\highlight}[1]{{\;\Strut\underline{#1}}\;}

\newcommand{\hohc}{\textsl{hohc}\xspace}
\newcommand{\fohh}{\textsl{fohh}\xspace}

\newcommand{\backw}{\mathrel{\texttt{:-}}}

\newcommand{\tm}{\hbox{\textsl{tm}}}
\newcommand{\ty}{\hbox{\textsl{ty}}}
\newcommand{\arrow}{\hbox{\textsl{arrow}}}
\newcommand{\app}{\hbox{\textsl{app}}}
\newcommand{\abs}{\hbox{\textsl{abs}}}
\newcommand{\toggle}[1]{\hbox{\textsl{toggle}$\;#1$}}
\newcommand{\swon}{\hbox{\textsl{sw\;on}}}
\newcommand{\swoff}{\hbox{\textsl{sw\;off}}}

\newcommand{\Lra}{\Longrightarrow}
\newcommand{\itm}[1]{\hbox{\textsl{item}}(#1)}

\newcommand{\fib}[1]{\hbox{\sl fib}\;#1}

\newcommand{\quest}{\mathord{?}}

\newcommand{\nbang}[1]{\hbox{$\bang^{#1}$}}
\newcommand{\etal}{\emph{et al.}}

\newcommand{\R}[2]{r\;#1\;#2}
\newcommand{\subst}[3]{[#1/#2]#3}
\newcommand{\tvar}{T}

\title{A Survey of the Proof-Theoretic Foundations of Logic Programming}
\author[Dale Miller]{Dale Miller\\
        Inria-Saclay \& LIX, Ecole Polytechnique, Palaiseau, France}

\date{\today}

\begin{document}
\label{firstpage}
\maketitle

\begin{center}
  {\bf Draft:} \today
\end{center}
\bigskip

\begin{abstract}
Several formal systems, such as resolution and minimal model
semantics, provide a framework for logic programming.  In this paper,
we will survey the use of \emph{structural proof theory} as an
alternative foundation.  Researchers have been using this foundation
for the past 35 years to elevate logic programming from its roots in
first-order classical logic into higher-order versions of 
intuitionistic and linear logic.  These more expressive logic
programming languages allow for capturing stateful computations and
rich forms of abstractions, including higher-order programming,
modularity, and abstract data types.  Term-level bindings are another
kind of abstraction, and these are given an elegant and direct
treatment within both proof theory and these extended logic
programming languages.  Logic programming has also inspired
new results in proof theory, such as those involving polarity and
focused proofs.  These recent results provide a high-level
means for presenting the differences between forward-chaining and
backward-chaining style inferences.  Anchoring logic programming in
proof theory has also helped identify its connections and differences
with functional programming, deductive databases, and model checking.
[To appear in {\em Theory and Practice of Logic Programming (TPLP)}.] 
\end{abstract}

\section{Introduction}
\label{sec:intro}

There are two broad approaches to relating logic with computational
systems \cite{miller06ijcar}.  On the one hand, there is the
\emph{computation-as-model} approach in which computations determine
models represented via mathematical structures containing such items
as nodes, transitions, and state.  Logic is used in an external sense
to make statements \emph{about} those structures.  That is,
computations are models, and logical expressions are evaluated over
such models.  Intensional operators, such as the modal operators of
temporal and dynamic logics or the triples in Hoare logic, are often
employed to express propositions about the state change.  This use of
logic to represent and reason about computation is probably the oldest
and most successful use of logic with computation.

On the other hand, the \emph{computation-as-deduction} approach uses
pieces of logic's syntax (\eg, types, terms, formulas, and proofs)
directly as elements of the specified computation.  There are two
different approaches to modeling computation in this much more
rarefied setting depending on how they use \emph{proofs}.  The
\emph{proof normalization} approach views the state of a computation
as a proof term and the process of computing as normalization (via
$\beta$-reduction or cut-elimination).  This approach to computing is
based on the \emph{Curry-Howard
correspondence}~\cite{curry34pnas,howard80,soerensen98book} and can
provide a theoretical framework for functional
programming~\cite{martinlof82}.  The \emph{proof search} approach
views the state of a computation as a \emph{sequent} (a particular
structured collection of formulas) and the process of computing as the
search for a proof of a sequent: the changes that take place in
sequents capture the dynamics of computation.  In the broadest sense,
proof search can be a foundation for interactive and automatic
theorem proving, model checking, and logic programming.  This paper
shall survey how the proof search interpretation of the sequent
calculus has been used to give a foundation to logic programming.

Unifying the two most foundational perspectives of logic---model
theory and proof theory---was the goal of some of the earliest work on
the foundations of logic programming.  However, these two perspectives
on logic have their own concerns and internal structure and results.
As a result, divergence appeared when these two perspectives were used
to motivate new designs and theories about programming with logic.
Taking models as primary, along with the direct treatment of negation
available in model theory, has led to new logic programming
languages, such as the \emph{answer set programming} approach to
declarative programming \cite{lifschitz08aaai,brewka11cacm} (see also
Section~\ref{ssec:advances}).  These developments have led to new
applications of logic in subjects such as databases, default
reasoning, planning, and constraint solving.  In this paper, we
survey, instead, the development of new logic programming language
designs and theories where proof theory is taken as primary.  Some
application areas of these designs have been type systems, proof
assistants, proof checking, and the specification of operational
semantics.

Symbolic logic is an appealing place to define a high-level
programming language for several reasons.  First, it is a well-studied
and mature formal language.  As a result, it has rich properties that
enable manipulating and transforming its syntax in meaning-preserving
ways.  Such manipulations include substitution into quantified
expressions, the unfolding of recursive definitions, and conversion to
normal forms (such as conjunctive normal form or negation normal
form).  Second, logics generally have multiple ways to look at what a
theorem is.  For example, soundness and completeness results allow us
to identify theorems as those formulas that have a proof and are true
in all models.  Finally, even for logics where model-theoretic
approaches are less commonly used, such as linear
logic~\cite{girard87tcs}, other deep principles, for example,
cut-elimination, are available.

Given that we choose to work with symbolic logic, how should we
connect logic with logic programming?  Clearly, the logical
foundation of Prolog---\emph{first-order Horn clauses}---should be
taken as an example of logic programming.  Nevertheless, the notion of
\emph{proof search} is a broad term, including, for example,
interactive and automated theorem provers where considerable
cleverness is needed to discover lemmas and inductive invariants.
Obviously, including the discovery of lemmas and invariants should not
be expected of an interpreter or compiler of a logic program.  Thus,
it seems necessary to draw a line between proof search in full logic
and some simpler, automatable subset of logic.

\section{The need for more expressive logic programming}
\label{sec:problems}

Horn clauses are formulas of the form
\begin{equation}
  \forall x_1\ldots\forall x_n [A_1 \land\ldots\land A_m \imp A_0]
  \qquad (n,m \ge 0).
\end{equation}
Here, the symbol $A$ (with or without subscripts and superscripts) is
used as a syntactic variable ranging over atomic formulas.
Notice that this formula can also be written without conjunctions as 
\begin{equation*}
  \forall x_1\ldots\forall x_n [A_1 \imp\ldots\imp A_m \imp A_0],
\end{equation*}
where the bracketing of implications is to the right.
In both cases, if $m=0$ then we do not write the implication.
A simple generalization of Horn clauses can be given by the following
grammar-like description of two classes of formulas using the
syntactic variables $G$ (for \emph{goal} formulas) and $D$ (for
\emph{definite} formulas).
\begin{align}
G & := A \sep \truth \sep G\land G \sep \false \sep G\lor G \sep 
              \exists x. G\label{hc:g}\\
D & := A\sep G\imp D  \sep \truth \sep D \land D \sep 
              \forall x. D\label{hc:d}
\end{align}
Here, $G$-formulas are freely generated from atomic formulas, $\truth$
(true), $\false$ (false), conjunction, disjunction, and existential
quantifiers.
A $D$-formula is a generalization of Horn clauses and these are such
that any subformula occurrence to the left of an implication is a
$G$-formula.
Using simple equivalences (which hold in classical and intuitionistic
logics), it is easy to show that a 
$D$-formula is logically equivalent to a conjunction of formulas
that are of the form $(1)$ above.\footnote{Throughout this paper, the
equivalence (in classical, 
intuitionistic, or linear logic) of two formulas $B$
and $C$ means that the two entailments $B\vdash C$ and $C\vdash B$ are
provable (in the respective logic).}

While the logic programs that can be written using first-order Horn
clauses are Turing complete~\cite{tarnlund77}, programming languages,
such as Prolog, based on Horn clauses have various weaknesses that
have been pointed out in the literature.  A list of some of these
shortcomings is below.

\begin{description}
  \item[Constraints:] The usual approach to data structures in Prolog
    encodes them as first-order terms using uninterpreted symbols.
    Occasionally, certain domains contain values that are much better
    handled by special-purpose algorithms instead of unification and
    strict syntactic equality.  Constraint logic programming
    \cite{jaffar87popl} is a general framework for organizing the
    treatment of such domains.

  \item[Negation-as-failure:] The simplest theories of Horn clauses do
    not include negation.  Different versions of negation, such as
    negation-as-failure~\cite{clark78}, have been added to most
    versions of Prolog.

  \item[Control of search:] Prolog implements depth-first search,
    which provides a natural procedural interpretation of many Horn
    clause specifications while providing expensive or non-terminating
    interpretations for other specifications.  Prolog has evolved
    several control mechanisms, such as \texttt{!} (cut), ancestor
    checking, and tabled deduction. 

  \item[Side-effects:] For a specification language to become a
    programming language, it seems necessary to accommodate primitives
    for side-effects and communications with other components of
    modern computer systems.  Primitives have been added to Prolog to
    allow side effects (e.g., \texttt{assert} and \texttt{retract}) 
    and input and output.

  \item[Abstraction mechanisms:] The logic behind Prolog does not
    directly support modern notions of abstractions, such as modules,
    abstract data types, higher-order programming, and binding
    structures.  Various extensions to Prolog addressing
    modular programming have been developed~\cite{iso00prolog2} and
    incorporated into most modern implementations of Prolog.

\end{description}

As this list shows, the development of programming language features
on top of Horn clauses has resulted in adding more to an exciting but
weak core logic setting.  The work that this survey explores takes a
different perspective to logic programming language
design.  Instead of working with a simple and weak foundation, proof
theory has been used to imagine large and more expressive logical
foundations, even going as far as adding higher-order quantification
and linear logic connectives.  Given the generality of such a large
framework, it is doubtful that the entire framework can be
effectively implemented.  However, the purpose of such imagining is
not to provide the foundations of a single, grand, practical logic
programming language but rather to develop a framework in which many
different sublanguages can be extracted (only one of which consists of Horn
clauses).  Such sublanguages would inherit some properties of the
larger framework, but their more narrow focus might allow for practical
implementations.  By way of analogy, consider the problem of building
parsers.  Context-free grammars (CFG) provide an important framework
for declaratively describing the structure of some languages.  Since
that framework is flexible and high-level, general-purpose parsers are
expensive: for example, the Earley parser has $O(n^3)$ complexity cost
for strings of length $n$~\cite{earley70cacm}.  Since this complexity
is too high for use in, say, compilers, many subsets of the general
CFG framework have been developed, such as the $LR(k)$ and $LALR(k)$
grammars, which describe fewer languages but have parsers with better
time and space complexity~\cite{aho07book}.  As we shall note in
Section~\ref{sec:prospects}, several subsets of the most general,
abstract logic programming framework have been identified and
implemented in different application areas.

\section{Some formal frameworks for logic programming}
\label{sec:frameworks}
 
A good formal framework for logic programming should satisfy some
properties, such as those listed here.
\begin{enumerate}

\item It should provide multiple and broad avenues for reasoning about
  logic programs.  We do not need new Turing machines because
  we do not need more specification languages that obviously
  compute but which do not come with support for addressing the
  correctness of specifications.

\item It should allow for the positioning of the logic programming
  paradigm among other programming and specification paradigms.

\item It should provide for a range of possible designs, leading to
  logic programming languages that go beyond the one acknowledged
  example based on Horn clauses.  Hopefully, these new designs would
  address some of the shortcomings outlined in the previous section.

\end{enumerate}

The following section focuses on the use of \emph{structural proof
  theory} as a foundation for logic programming.  The rest of this
section describes three other popular approaches to the formal
foundations for logic programming.

\subsection{Resolution}
\label{ssec:resol}

Following Robinson's introduction of the resolution refutation method
for automating first-order logic~\cite{robinson65jacm}, several
researchers developed strategies to tame the search for refutations.
One of these strategies is linear resolution, which was developed
independently by Loveland~\citeyear{loveland70lnm},
Luckham~\citeyear{luckham70lnm}, and Zamov and Sharonov
\citeyear{zamov71scmml}.  After the first Prolog system was developed
at the University of Aix-Marseilles in 1972
\cite{colmerauer93sigplan}, Kowalski \citeyear{kowalski74ip}
formalized its operational behavior as linear resolution.  Later, Apt
and van Emden~\citeyear{apt82jacm} named this particular style of
resolution \emph{SLD-resolution} (``Selective Linear Definite clause
resolution'') and proved it to be complete when restricted to Horn
clauses.

As a framework, resolution has been used to provide a treatment of
some extensions of Prolog.  For example, Clark \citeyear{clark78}
introduced the \emph{if-and-only-if} completion of Horn clauses.  He
showed how the failure of an exhaustive and finite search for an
SLD-resolution refutation could be used to justify a proof of a negated goal.
Clark's extended refutation procedure, now called SLDNF, received
various descriptions and correctness proofs (see, for example,
the papers by Apt and Bol \citeyear{apt94jlp} and by Apt and
Doets~\citeyear{apt94jlpb}) and has been used to extend logic 
programming to include all of first-order classical logic
\cite{lloyd84jlp}.  Resolution also allows for a simple approach to
the treatment of constraints and their flexible scheduling
\cite{huet73ijcai}.  Minker and Rajasekar \citeyear{minker90jlp}
specialized resolution to serve as a proof strategy for disjunctive
logic programming.  Loveland's Near-Horn Prolog
\cite{loveland87iclp,loveland91cl} was also described using
resolution, although it was eventually given a description using
sequent calculus proofs~\cite{nadathur95lics,nadathur98handbook2} in
the style we shall see in Section~\ref{sec:intuitionistic}.

Although resolution refutations had some successes as a framework for
logic programming, this framework has been problematic for at least
two reasons.  First, it generally relies on normal forms, such as
conjunctive normal form, negation normal form, and Skolem
normal form.  Such normal forms are not generally available outside of
first-order classical logic: in particular, these normal forms are not
sound for intuitionistic logic.  Secondly, resolution is pedagogically
flawed since it forces the attempt to \emph{prove} the goal $G$
from the program $\Pscr$ into the attempt to \emph{refute} the set of
formulas $\Pscr\cup\{\neg G\}$; switching from proving to refuting is
unfortunate, unintuitive, and, as we shall see, unnecessary.  As the
author has argued elsewhere \citeyear{miller21pt}, the use of
Skolemization to simplify the structure of quantifiers in formulas
appears to be the dominant reason for early automated theorem proving
systems to rely on building refutations instead of proofs.  Since
structural proof theory provides an alternative to Skolemization, that
framework can rely on proving instead of refuting.

\subsection{Model theory}
\label{ssec:models}

Given the success of denotational semantics to provide a
mathematically precise notion of meaning for various programming
languages~\cite{scott70,stoy77} and given that model theory for
first-order classical logic was a well-developed topic before the
advent of logic programming, it was natural to consider using
model theory as a semantics for logic programming.

Apt, van Emden, and Kowalski provided the first steps to building such
a semantics for logic programming.   They connected
SLD-resolution to fixed-point operators on models represented by sets
of atomic formulas.  In particular, the least-fixed point model
semantics was shown to characterize provable atomic formulas, while
negation-as-failure was shown to relate to the greatest-fixed point
model \cite{apt82jacm,emden76jacm}.  Model theory has also been used
to provide various formal definitions of negation-as-failure,
including \emph{well-founded semantics} \cite{vangelder91jacm} and
\emph{stable models} \cite{gelfond88iclp}.

Model theory can sometimes be used to provide an equivalent
perspective on provability.  In particular, the familiar soundness and
completeness theorems state that provable statements are exactly the
valid statements.  Such a result can convince us that a given proof
system is not, in fact, ad hoc, inconsistent, or missing inferences or
axioms.  Such confidence indeed arises from the earliest completeness
theorems, such as the ones given by G\"odel \citeyear{goedel30mfm} and
Henkin \citeyear{henkin49jsl}.  Today, however, experts in model
theory and category theory have sufficient ``muscle'' so that they can
build complicated and ad hoc semantic domains.  As a result, soundness
and completeness theorems are not as compelling as they once were.
Fortunately, proof theory comes with its own principles, such as, for
example, the cut-elimination theorem, which helps to rule out ad hoc
and inconsistent inference principles.

\subsection{Operational semantics}
\label{sec:2-1-3}

Semantics can be given for logic programming by providing a
mathematical description of the language's behavior.  One such
approach has been to use \emph{abstract machines}, such as the Warren
Abstract Machine WAM \cite{warren83tr,aitkaci91wam}, to describe the
behavior of logic programs.  Such machines can be taken as formal
models when they are given a formal specification
\cite{borgere95lpfmpa}.

A few high-level and formal specifications of parts of Prolog's
operational semantics have been developed starting a couple of decades
ago and using different techniques: e.g., Andrews
\citeyear{andrews97tcs} used a combination of a multi-valued logic and
a transition system, Li \citeyear{li94esop} used the $\pi$-calculus,
and B\"orger and Rosenzweig \citeyear{borger95scp} used evolving
algebras.  Such approaches to specification have the advantage that
they can describe the actual behavior of Prolog implementations when
they need to deal with features such as the cut \texttt{!} control
operator and the \texttt{assert} and \texttt{retract} predicates.
Such features are difficult or impossible to address using resolution
refutations or model theory.

Since these specification styles are formal, any attempt to reason
about them also certainly requires using proof assistants.  These
specifications are used to address the question ``How do we implement
a language?'' and not the more general question ``What language should
we implement?''  While the former question is important, we shall
focus on a framework that addresses the latter question.

\section{The trajectory of proof theory investigations}
\label{sec:trajectory}

The term ``proof theory'' is often used in the logic programming
literature to refer to some characterization of provability $\vdash$
in contrast to validity $\models$.  However, in many texts,
provability is characterized indirectly using resolution refutations.
In this paper, the term ``proof theory'' is used exclusively to refer
to the systems and methods introduced by Gentzen in his famous paper
\citeyear{gentzen35}.  In that paper, Gentzen introduced both natural
deduction and the sequent calculus and proved the cut-elimination
theorem for classical and intuitionistic logics.  Gentzen's proof
systems have been applied in many different settings during the past
several decades.  In mathematics, they have been used to prove the
consistency of various logical and arithmetic systems
\cite{gentzen35}; in logic, they have been used to define various
modal logics \cite{ono98msj,wansing02handbook}; in linguistics, they
have been used to describe the structure of sentences~\cite{lambek58};
and in computational logic, they have been used to provide the formal
setting for discussing both computing via proof normalization and
proof search (see Section~\ref{sec:intro}).  We shall be particularly
interested in using Gentzen's proof systems to analyze the syntax and
structure of proofs themselves.  With this emphasis, this topic is
often called \emph{structural proof theory}.  Good background material
on this topic can be found in the papers by
Gallier~\citeyear{gallier86}, Girard et al.~\citeyear{girard89book},
Buss~\citeyear{buss98handbook}, and Negri and von
Plato~\citeyear{negri01spt}.  Since the $\lambda$-calculus will also
be associated with our discussion of proof theory, the reader
unfamiliar with the basics of the $\lambda$-calculus can find good
background material in the work of Barendregt~\citeyear{barendregt84},
Huet~\citeyear{huet75tcs}, Miller and
Nadathur~\citeyear{miller12proghol}, and Barendregt et
al.~\citeyear{barendregt13book}.

The following list of key applications of structural proof theory to
logic programming helps to provide an outline of the rest of this
survey.
\begin{enumerate}

\item In Section~\ref{sec:gdp}, we position logic programming within
  the sequent calculus instead of resolution and then describe the
  nature of \emph{goal-directed search} and \emph{backward chaining}
  using sequent calculus inference rules.

\item Given that sequent calculus proof systems were known for
  first-order and higher-order classical and intuitionistic logics,
  the first proof-theoretic extensions of logic programming were
  investigated in these logics.  For example, developing proof search
  results within higher-order intuitionistic logic provided logic
  programming with various forms of abstractions,
  including higher-order programming, modules,
  and abstract data types (see Sections~\ref{sec:ho}
  and~\ref{sec:modules}).  The sequent calculus also enables a new
  treatment of binding structures (within terms, formulas, and
  proofs): this treatment is described in Section~\ref{sec:mobility}.

\item The appearance of linear logic provides new and sometimes
  surprising avenues for extending logic programming to settings
  involving stateful and concurrent computations (see
  Section~\ref{sec:llp}).

\item The proof theory of linear logic introduced the notions of
  \emph{polarity} and \emph{focused proofs} (see
  Section~\ref{sec:focusing}).  When these notions are applied to
  logic programming, they allow for extending the notion of
  goal-directed proof.  These notions also provide an elegant
  description of both forward-chaining and backward-chaining
  inference.
\end{enumerate}

\section{Intuitionistic logic and proof search}
\label{sec:intuitionistic}

In the 1980s, there were some early attempts to use various proof
systems as frameworks for logic programming based on extended versions
of Horn clauses.  For example, Hagiya and
Sakurai~\citeyear{hagiya84ngc} used Martin-L\"of's theory of iterated
inductive definitions~\citeyear{martin-lof71sls} to describe Horn
clause reasoning and negation-as-failure.  There were also several
attempts to extend Prolog to full first-order logic.  In particular,
Bowen \citeyear{bowen82mi} described how sequent calculus and
unification could be merged; Haridi and Sahlin~\citeyear{haridi83lpw}
described an implemented proof system using natural deduction; and
Cellucci~\citeyear{cellucci87} proposed using tableaux proof systems for
the specification of logic programming.

In the second half of the 1980s, several researchers discovered
roles for intuitionistic logic within computational logic that were not
directly related to the Curry-Howard correspondence (briefly described in
Section~\ref{sec:intro}).  Instead, these 
roles supported the proof-search paradigm.  These discoveries, listed
below, were made nearly simultaneously and largely independently.

\begin{itemize}
\item Gabbay and Reyle developed N-Prolog, an extension to Prolog with
  hypothetical goals \cite{gabbay84,gabbay85jlp}.

\item The \lP logic programming language by Nadathur and the author
  \citeyear{nadathur88iclp,miller12proghol} lifted the Prolog language
  to higher-order intuitionistic logic.  This logic provided
  hypothetical and generic reasoning as well as higher-order
  programming for logic programming \cite{miller91apal}.

\item McCarty \citeyear{mccarty88a,mccarty88b} used Kripke model
  semantics of intuitionistic logic to study an extension of logic
  programming that supported hypothetical reasoning.

\item Paulson \citeyear{paulson89jar} used natural deduction and
  intuitionistic logic to provide a framework for the generic theorem
  prover at the core of the Isabelle prover.  Some design and
  implementation issues in that prover are closely related to design
  and implementation issues in the \lP system.

\item Halln\"{a}s and Schroeder-Heister
  \citeyear{hallnas90jlc,hallnas91jlc} also explored a logic
  programming interpretation of hypothetical reasoning using the proof
  theory of intuitionistic logic.

\item Mints and Tyugu \citeyear{mints90jlp} used propositional
  intuitionistic logic to design and automate their PRIZ programming
  system.
\end{itemize}

Also, during this period, the dependently typed $\lambda$-calculus LF
\cite{harper93jacm} was proposed as a framework for describing proof
systems for intuitionistic logic: it was also given a \lP-inspired
implementation within the Elf system~\cite{pfenning89lics}.

In the \emph{proof search} setting, the successful completion of a
(non-deterministic) computation is encoded by a cut-free proof.  Here,
proof normalization and cut elimination are not part of the
computation engine but instead can be involved in reasoning about
computation.  See Section~\ref{ssec:reasoning} for a discussion about
proof-theoretic methods for reasoning about logic programs.

\subsection{Provability via the Sequent calculus}
\label{sec:lj}

While we assume that the reader has some familiarity with the sequent
calculus, we review some basic concepts.  Formally, a \emph{sequent}
is a pair of multisets of formulas, written as $\Seq{\Gamma}{\Delta}$,
and we speak of a formula occurrence in $\Gamma$ as being on the
left-hand side and a formula occurrence in $\Delta$ as being on the
right-hand side of that sequent.  Gentzen's proof system for classical
logic, called \LK~\cite{gentzen35}, allows any number of formulas in
$\Delta$, whereas his proof system for intuitionistic logic, called
\LJ, requires $\Delta$ to contain at most one formula.  Otherwise, proofs in
intuitionistic and classical logics use the same set of inference
rules.

Inference rules that deal directly with logical connectives are called
\emph{introduction rules} and are used to introduce logical
connectives into the right or left sides of a sequent.  The following
three inference rules are used to introduce the conjunction,
disjunction, and universal quantifier into the left-hand sides of
sequents.
\[
  \infer[\wedge L]{\Seq{\Gamma, B\wedge C}{E}}{\Seq{\Gamma, B, C}{E}}
  \qquad\quad
  \infer[\vee L]{\Seq{\Gamma, B\vee C}{E}}
                {\Seq{\Gamma, B}{E} \qquad \Seq{\Gamma, C}{E}}
  \qquad\quad
  \infer[\forall L]{\Seq{\Gamma, \forall x. B}{E}}{\Seq{\Gamma, \subst{t}{x}{B}}{E}}
\]
The $\wedge L$ rule says that one way to prove that $E$ follows
from $B\wedge C$ and $\Gamma$ is to prove that $E$ follows from $B$
and $C$ and $\Gamma$.  The $\vee L$ rule is the sequent calculus
version of the \emph{rule of cases}\/: one way to prove that $E$
follows from $B\lor C$ and $\Gamma$ is to prove that $E$ follows from
$B$ and $\Gamma$ (the first case) and that $E$ follows from $C$ and
$\Gamma$ (the second case).  The $\forall L$ rule says that one way to
prove that $E$ follows from $\forall x. B$ and $\Gamma$ is to prove
that $E$ follows from $\subst{t}{x}{B}$ and $\Gamma$, where $t$ is some term
and $\subst{t}{x}{B}$ is the (capture avoiding) substitution of $t$ for $x$ in
$B$.  Figure~\ref{fig:lj} contains introduction rules for the
implication and the universal quantifier.

At least three significant problems with the sequent calculus
translate into difficulties using it as a foundation for logic
programming.  Unlike resolution refutations, the sequent calculus is
not equipped with unification, which is recognized as an essential
operation in logic programming.  For example, in the $\forall L$ rule
above, the substitution instance $t$ for $\forall x.B$
must be chosen when this rule is applied, even though the exact nature
of that term may not be known in detail until much later in the
search for a proof.  We set this problem aside until
Section~\ref{sec:mobility}.

A second serious problem with applying the sequent calculus to logic
programming is that its proofs are formless, low-level, and painful to
use directly.  To illustrate this problem, consider the situation
where $A$ is an atomic formula and $\Gamma$ is a multiset of 998
non-atomic formulas, and where we wish to find a proof of the
sequent $$\Seq{\Gamma, B_1\vee B_2, C_1\wedge C_2}{A}.$$
There can be
1000 choices of left introduction rules to attempt in order to prove this
sequent.\footnote{We shall use the left-hand side of a sequent to store a logic
program.  A logic program with 1000 formulas (clauses) is a 
small-to-medium-sized program.}
Once one of those choices is made, it is likely that that
rule yields at least one premise that again has about 1000 non-atomic
formulas on the left.  For example, first applying $\vee L$ and then
applying $\wedge L$ on each premise can yield
\[\infer[\vee L.]{\Seq{\Gamma, B_1\vee B_2, C_1\wedge C_2}{A}}
  {\infer[\wedge L]{\Seq{\Gamma, B_1, C_1\wedge C_2}{A}}{\Seq{\Gamma,
        B_1, C_1, C_2}{A}} &
   \infer[\wedge L]{\Seq{\Gamma, B_2, C_1\wedge C_2}{A}}{\Seq{\Gamma,
        B_2, C_1, C_2}{A}}
}
\]
This tiny proof fragment is roughly one of about a million choices.
Equally unfortunate is what happens if the search for a proof fails to find a
proof of the left premise.  The proof procedure could then choose to
do these two inference rules in the opposite order, namely giving rise
to the proof fragment
\[\infer[\wedge L.]{\Seq{\Gamma, B_1\vee B_2, C_1\wedge C_2}{A}}
  {\infer[\vee  L]{\Seq{\Gamma, B_1\vee B_2, C_1, C_2}{A}}
                  {\Seq{\Gamma, B_1, C_1, C_2}{A} \qquad
                   \Seq{\Gamma, B_2, C_1, C_2}{A}}}
\]
However, this \emph{permutation} of inference rules yields the same
premises.  As a result, proof search will again fail on the left
branch.  Clearly, switching the order of these rules is not important
for completeness.  Any high-level structure that sequent calculus
proofs might contain needs to be pulled out by extensive inference
rule permutation arguments.  Such high-level structure in proofs will
be more apparent when we upgrade sequent calculus proofs to
\emph{focused} proofs in Section~\ref{sec:focusing}.

The third serious problem with applying the sequent calculus
to logic programming is that its inference rules are too tiny and not
the right inference rules in many settings.  For example, consider a
multiset of formulas $\Gamma$ that contains the following two Horn
clauses.
\begin{align*}
\forall x\forall y [\adj x y \imp \pth x y]\qquad\quad\\
\forall x\forall y \forall z [\pth x z \imp \pth z y\imp \pth x y]
\end{align*}
The effect of using these formulas in a proof can naturally be 
viewed as describing inference rules directly.  For
example, the backward-chaining interpretation of these formulas seems
best captured using the following pair of rules:
\[
  \infer{\Seq{\Gamma}{\pth x y}}{\Seq{\Gamma}{\adj x y}}
  \qquad\quad
  \infer[.]{\Seq{\Gamma}{\pth x y}}{\Seq{\Gamma}{\pth x z}\quad \Seq{\Gamma}{\pth z y}}
\]
The forward-chaining interpretation of these formulas seems best
captured using the following pair of rules:
\[
  \infer{\Seq{\Gamma,\adj x y}{A}}{\Seq{\Gamma, \adj x y, \pth x y}{A}}
  \qquad
  \infer[.]{\Seq{\Gamma,\pth x z,\pth z y}{A}}{\Seq{\Gamma,\pth x z,\pth z y,\pth x y}{A}}
\]
Note that none of these rules explicitly contain occurrences of
logical connectives.  When we deal with \emph{polarization} and
\emph{focused} proofs in Section~\ref{sec:focusing}, we will show how
to construct these inference rules from Horn clauses and how
polarization selects between rules following the backward-chaining or
forward-chaining discipline.

\subsection{Goal-directed proofs}
\label{sec:gdp}

\begin{figure}
\begin{flushleft}
  \textsc{Structural rules}
\end{flushleft}
\[
  \infer[\contrL]{\Seq{\Gamma,B}{E}}{\Seq{\Gamma,B,B}{E}}
  \qquad\qquad
  \infer[\weakL ]{\Seq{\Gamma,B}{E}}{\Seq{\Gamma}{E}}
\]
\begin{flushleft}
  \textsc{Identity rules}
\end{flushleft}
\[
  \infer[\init]{\Seq{B}{B}}{}
  \qquad\qquad
  \infer[\cut]{\Seq{\Gamma_1,\Gamma_2}{E}}
              {\Seq{\Gamma_1}{B}\qquad\Seq{\Gamma_2,B}{E}}
\]
\begin{flushleft}
  \textsc{Introduction rules}
\end{flushleft}
\[
  \infer[\impL]
        {\Seq{\Gamma_1,\Gamma_2,B_1 \supset B_2}{E}}
        {\Seq{\Gamma_1}{B_1}\qquad \Seq{\Gamma_2,B_2}{E}}
  \qquad
  \infer[\impR]{\Seq{\Gamma}{B_1 \supset B_2}}{\Seq{\Gamma,B_1}{B_2}}
\]
\[
  \qquad\infer[\forallL]{\Seq{\Gamma,\forall x.B}{E}}{\Seq{\Gamma,\subst{t}{x}{B}}{E}}
  \qquad\qquad\quad
  \infer[\forallR]
        {\Seq{\Gamma}{\forall x.B}}{\Seq{\Gamma}{\subst{y}{x}{B}}}
\]

\caption{The subset of Gentzen's \LJ proof system that applies to only
  $\supset$ and $\forall$.  In the $\forallR$ rule, the variable $y$ is
  not free in any formula in the conclusion of that rule.}
\label{fig:lj}
\end{figure}

Throughout this survey, we shall see several different sequent calculi
presented as a collection of inference rules.  To simplify the
presentation and comparison of such systems, we shall usually restrict
our attention to formulas containing just the logical connectives for
universal quantification and implication.  For example,
Figure~\ref{fig:lj} contains the subset of Gentzen's \LJ proof
system~\cite{gentzen35} that applies to only the logical connectives
$\supset$ and $\forall$.  Proof systems in the literature (for example,
\cite{gentzen35}) usually contain more 
logical connectives (\eg, disjunction, conjunction, and existential
quantifiers).  The variable $y$ used in the $\forallR$ rule is called
the \emph{eigenvariable} for that rule.

An early application of sequent calculus to logic programming was the
development of the technical term \emph{uniform proof} to
capture the notion of \emph{goal-directed search} \cite{miller91apal}.
In particular, the sequent $\Seq{\Pscr}{G}$ describes the
obligation to prove the \emph{goal formula} $G$ from the \emph{(logic)
program} $\Pscr$.  

To formalize the fact that a proof attempt is goal-directed, we will
insist that whenever the goal formula is non-atomic (hence,
its top-level symbol is a logical connective), this sequent can only
be proved using a right introduction rule.  Even if the left-hand side
$\Pscr$ contains 1000 non-atomic formulas, a goal-directed proof must
ignore the possibility of introducing those formulas and only allow
the right-hand formula to be selected.  Only when the top-level
symbol of the goal formula is non-logical (that is, it is a predicate
symbol) is the proof attempt permitted to consider the left-hand
side.  Such sequent calculus proofs were called \emph{uniform proofs}
\cite{miller91apal}.  In general, a uniform proof is divided into two
phases.  One phase involves a sequence of right-introduction rules
that performs goal reduction. The other phase---the
\emph{backward-chaining phase}---selects a formula from the left-hand
side $\Pscr$ and performs a sequence of left-introduction rules
derived from that one formula.

To illustrate a backward-chaining phase, consider the following proof
fragment.  Here, $\Pscr$ is a multiset of formulas that includes the
two Horn clauses in the previous section that describe the
\textsl{path} predicate.
\[
\infer[\contrL]
      {\Pscr\vdash \pth{a}{c}}{
\infer[\forall L\times 3]
      {\Pscr,\highlight{\forall x\forall y\forall z(\pth{x}{y}\supset \pth{y}{z}\supset \pth{x}{z})}
                    \vdash \pth{a}{c}}{
\infer[\contrL]{\Pscr,\highlight{ \pth{a}{b}\supset \pth{b}{c}\supset \pth{a}{c}}\vdash \pth{a}{c}}{
\infer[\supset L]{\Pscr,\Pscr,\highlight{ \pth{a}{b}\supset \pth{b}{c}\supset \pth{a}{c}}\vdash \pth{a}{c}}{
   \deduce{\Pscr\vdash \pth{a}{b}}{} & 
   \infer[\supset L]{\Pscr,\highlight{ \pth{b}{c}\supset \pth{a}{c}}\vdash \pth{a}{c}}{
      \deduce{\Pscr\vdash \pth{b}{c}}{} &
      \infer[\init]{\highlight{ \pth{a}{c}}\vdash \pth{a}{c}}{}}}}}}
\]
Here, a formula on the left is highlighted by underlining it.  The
backward-chaining phase has four important features.  First, it is
invoked only if the goal on the right is atomic.  Second, only the
highlighted formula is the site of a left-introduction rule.  Third,
if the highlighted formula is atomic, then the sequent in which it
occurs must be the conclusion of the $\init$ rule: i.e., the
highlighted formula and the goal must be 
equal.  Fourth and finally, the contraction rule is responsible for
selecting a formula on which to focus.  Note that during all steps in
building this phase, the contraction rule may have to make many
choices: once that choice is taken, there is no longer any choice as
to which left-introduction rule gets applied.  If we erase all
sequents in the fragment above containing an underlined formula, then
the result is exactly one of the backward-chaining inference rules
from the previous section.

It is now possible to put these various notions together and define
an \emph{abstract logic programming language} as a
triple $\tup{\Dscr,\Gscr,\vdash}$ such that for all finite subsets
$\Pscr$ of $\Dscr$ and all formulas $G$ of $\Gscr$, $\Pscr\vdash G$ is
provable if and only if the sequent $\Seq{\Pscr}{G}$ has a uniform
proof.

Let $\Dscr_1$ and $\Gscr_1$ be collections of Horn clauses and goal
formulas as described by lines $(\ref{hc:g})$ and $(\ref{hc:d})$ in
Section~\ref{sec:problems}.  Using basic proof theory arguments, it is
easy to show that $\tup{\Dscr_1,\Gscr_1,\vdash_C}$ and
$\tup{\Dscr_1,\Gscr_1,\vdash_I}$ are both abstract logic programming
languages (here, $\vdash_C$ and $\vdash_I$ denote provability in
classical and intuitionistic logics, respectively).  Thus, Horn
clauses---using intuitionistic or classical logic---provide an example
of an abstract logic programming language.  In a sense, Horn clauses
form a setting that is so weak that it cannot distinguish between
classical and intuitionistic provability.

\subsection{Higher-order Horn clauses}
\label{sec:ho}

While the functional programming world has embraced higher-order
programming since its inception, logic programmers have often held such
programming style at an arm's length.  For example, D. H. D. Warren
\citeyear{warren82mi} argued that higher-order predicate
quantification could be translated away and, as a result, an explicit
higher-order extension to logic programming was not needed.
Similarly, HiLog~\cite{chen93jlp} added mild extensions to the syntax
of Prolog to accommodate some aspects of higher-order programming, but
HiLog was restricted to maintain the first-order aspects of the
underlying implementation of Prolog.

Although Church did not use structural proof theory to
introduce the higher-order logic he called the \emph{Simple Theory of
  Types} \citeyear{church40}, several proof-theoretic treatments of
the classical and intuitionistic versions of higher-order logic were
developed in the decades following its introduction
\cite{takeuti53jjm,takahashi67jmsj,girard71sls}.  When
Gentzen's notion of sequent calculus is used to describe the classical
and intuitionistic versions of Church's Simple Theory of Types, one
gets an elegant proof system for very expressive logics.  There was
also early work on implementing various aspects of theorem
proving in Church's logic, including unification \cite{huet75tcs},
resolution \cite{andrews71jsl,huet73ijcai}, and general theorem
proving \cite{andrews84ams,paulson89jar}.

Starting with that earlier work, Nadathur and the author worked on
trying for a genuine, higher-order logic generalization to logic
programming.  They defined a notion of \emph{higher-order Horn
clauses} (\hohc), proved that they formed an abstract logic
programming language, and described the design of an interpreter for
what was the basis of an early version of \lP
\cite{miller86iclp,nadathur87phd,nadathur90jacm}.  In this new logic
programming language, it is easy to write higher-order
programs, such as the following (using \lP syntax).
\begin{verbatim}
type foreach, forsome   (A -> o) -> list A -> o.
type mappred            (A -> B -> o) -> list A -> list B -> o.

foreach P [].
foreach P [X|L] :- P X, foreach P L.

forsome P [X|L] :- P X; forsome P L.

mappred P [] [].
mappred P [X|L] [Y|K] :- P X Y, mappred P L K.
\end{verbatim}
In the first two lines above, the types of three higher-order
predicates are declared.  These type expressions follow a convention
begun by Church \citeyear{church40} in which \verb+o+ is used to
denote the type of formulas.  Thus, a symbol of type \verb+nat -> o+
denotes a predicate of one argument of type \verb+nat+.  Capital letters
in type expressions denote type variables: thus, these type
declarations are polymorphically typed in a sense similar to, say,
ML~\cite{milner90sml,nadathur92types}.

Actually, \hohc contained more than is necessary to capture
higher-order relational programming.  For example, it includes
quantification over non-primitive and non-relational types as well as
the simply-typed $\lambda$-calculus with equality and unification
modulo $\alpha$, $\beta$, and $\eta$ conversions.  As a result, some
sophisticated computations on syntactic expressions containing
bindings are possible in \hohc, including, for example, program
transformations \cite{huet78} and natural language semantics
\cite{miller86acl}.  For those not interested in dealing with bindings
in term structures, Wadge \citeyear{wadge91ilps} and Bezem
\citeyear{bezem01csl} have developed restrictions to \hohc that seem
to capture what is needed for higher-order relational programming
(including those displayed above).

\subsection{Hypothetical goals and modular structures}
\label{sec:modules}

Consider the following definition for a larger class of definite
(program) clauses and goal formulas that extends the corresponding
definition for Horn clauses and their goals (given by lines
$(\ref{hc:g})$ and $(\ref{hc:d})$ in Section~\ref{sec:problems}) by
adding implication and universal quantification to goals.
\begin{align}
G & := A \sep \truth \sep G\land G \sep \false \sep G\lor G \sep 
       \exists x. G\sep D\imp G \sep \forall x. G\label{hh:g}\\
D & := A\sep G\imp D  \sep \truth \sep D \land D \sep \forall x. D\label{hh:d}
\end{align}
These definitions of $G$ and $D$-formulas are mutually recursive.  This
definition is large enough to contain the extended logic programming
systems that were mentioned at the beginning of Section~\ref{sec:intuitionistic}.
Compare the definition for $D$ above with the following definition of
\emph{Harrop} formulas \cite{harrop60jsl}:
\begin{align*}
H & := A\sep B\imp H \sep \truth \sep H \land H \sep \forall x. H,
\end{align*}
where the syntactic variable $B$ ranges over arbitrary
first-order formulas.  The $D$ formulas in $(\ref{hh:d})$ are such
that any negatively occurring subformula of a $D$ formula is such that
negatively occurring subformulas of them are again Harrop formulas.
Motivated by this observation, such $D$ formulas are called first-order
\emph{hereditary Harrop formulas} (\fohh) \cite{miller91apal}.  

Let $\Dscr_2$ and $\Gscr_2$ be collections of formulas described by
lines $(\ref{hh:g})$ and $(\ref{hh:d})$.  Using inductive arguments
with sequent calculus proofs, it can be shown that the triple
$\tup{\Dscr_2,\Gscr_2,\vdash_I}$ is an abstract logic programming
language.  Unlike the case with Horn clauses, the triple
$\tup{\Dscr_2,\Gscr_2,\vdash_C}$ is \emph{not} an abstract logic
programming language.  To see this, consider the goal formula $(p\imp
q)\vee p$, where $p$ and $q$ are two propositional constants.  
Note that this formula is classically provable since we have the
following classical equivalences (using only the associativity and
commutativity of $\vee$ and the equivalence $B\imp C\equiv \neg B\vee
C$):
\[
  (p\imp q)\vee p      \equiv 
  (\neg p\vee q)\vee p \equiv 
  q\vee (\neg p\vee p) \equiv 
  q\vee (p\imp p).
\]
Since the last of these formulas is true, they are all true
and classically provable.  However, the sequent $\Seq{}{(p\imp q)\vee
  p}$ does not have a uniform proof since uniformity requires that
this sequent is the conclusion of a right introduction of $\vee$ in
which case either $\Seq{}{p\imp q}$ or $\Seq{}{p}$ are provable: but,
of course, neither of these sequents are provable.

Goal formulas in \fohh allow \emph{hypothetical} goals: if we attempt
to find a uniform proof of the sequent $\Seq{\Pscr}{D\imp G}$ then
that attempt leads to attempting to prove $\Seq{\Pscr,D}{G}$.  Note
that the left-hand side of sequents can get larger as one searches for
a uniform proof.  For example, an attempt to find a uniform proof of
the sequent
\[
  \Seq{\Pscr}{(D_1\imp (G_1\wedge (D_2\imp G_2))) \wedge G_3\wedge
    (D_3\imp G_4)}
\]
must lead to attempts to prove the four sequents
\[
  \Seq{\Pscr,D_1}{G_1} \qquad \Seq{\Pscr,D_1, D_2}{G_2} \qquad 
  \Seq{\Pscr}{G_3} \qquad \Seq{\Pscr,D_3}{G_4}.
\]
Hence, the logic program (the left-hand side) for these subgoals can
all be different.  Such an observation has been used to design a
logic-based approach to modular programming within logic programming:
in particular, the goal $D\imp G$ can be operationally interpreted 
to mean that an individual goal can request that the program clauses
in $D$ are loaded before attempting to prove $G$ \cite{miller89jlp}.

The use of hypothetical goals to load code  available for a
certain scope is not supported by classical logic.  In particular, the
intuitionistic logic interpretation of the goal $(D \supset G) \lor H$
means that $D$ is available during the search for the proof of $G$ but
not $H$.  In classical logic, this scoping breaks down since,
classically, this formula is equivalent to $G \lor (D \supset H)$ and
to $(D \supset (G \lor H))$ (using the classical equivalences
mentioned above).  An overview of modularity proposals for logic
programming based on proof theory, Kripke semantics, modal operators,
and algebraic operators can be found in the survey by Bugliesi et
al.~\citeyear{bugliesi94jlp}. 

Several researchers have used uniform proofs to motivate new proof
procedures and new logic programming designs.  Nadathur
\citeyear{nadathur93jar} presented a proof procedure for hereditary
Harrop formulas in which the interplay of unification and
eigenvariables is explicitly treated.  Harland \citeyear{harland97cl} and
Nadathur \citeyear{nadathur98jlc,nadathur00tcs} have also provided new
proof systems for classical logic based on using uniform proofs.
Recently, such proofs have also been used to describe a coinductive
proof procedure for Horn clauses \cite{basold19esop}.  As we shall see
in Section~\ref{sec:llp}, uniform proofs have also been used to design
various \emph{linear logic programming languages}.

\section{Universal goals, binder mobility, and abstract data types}
\label{sec:mobility}

Logic programming based on hereditary Harrop formulas allows goals to
be universally quantified.  We now describe the operational
significance of that quantifier in logic programming.  Once again,
Gentzen's sequent calculus provides an elegant treatment for such
quantification using the notion of \emph{eigenvariable}.

\subsection{Eigenvariables as explicit bindings}

The $\forallR$ inference rule in Figure~\ref{fig:lj} replaces the
universally quantified binding for $x$ in its conclusion with a free
variable $y$ in its premise.  The $47^{th}$ epigram of Alan Perlis
\citeyear{perlis82sigplan} is worth repeating here: ``As Will Rogers
would have said, `There is no such thing as a free variable.' '' The
wisdom of this epigram is that free variables are, in fact, bound (or
declared) somewhere, and that place should be made explicit.  To this
end, consider replacing the sequent $\Seq{\Gamma}{E}$ with
$\Seqq{\Sigma}{\Gamma}{E}$, where $\Sigma$ is a list of distinct
variables that are considered bound over both $\Gamma$ and $E$.  Thus,
the eigenvariables for a sequent are explicitly bound over the
sequent.  The inference rules for the universal quantifier in
Figure~\ref{fig:lj} are then changed as follows.
\[
  \infer[\forallL]{\Seqq{\Sigma}{\Gamma,\forall x_\tau.B}{E}}
                  {\Sigma\vdash t:\tau\qquad\Seqq{\Sigma}{\Gamma,\subst{t}{x}{B}}{E}}
  \qquad\qquad
  \infer[\forallR]{\Seqq{\Sigma}{\Gamma}{\forall_\tau x.B}}
                  {\Seqq{\Sigma,y:\tau}{\Gamma}{\subst{y}{x}{B}}}
\]
In the $\forallR$ rule, the eigenvariable is now explicitly bound
within the sequent.  The $\forallL$ rule is also updated with the
premise $\Sigma\vdash t:\tau$ that ensures that the term $t$ is
built from only the eigenvariable in the context $\Sigma$.  If we
limit ourselves to a simple single-sorted first-order logic, then the
type variable $\tau$ is set to just that sort.  If we are
working with a multi-sorted logic, $\tau$ could range over the various
sorts.  Finally, if we are dealing with the \emph{elementary type
  theory} \cite{andrews74zmlgm} subset of Church's Simple Theory of
Types \cite{church40}, then $\tau$ could range over all simple types,
including higher-order types.

The other inference rules are also given the $\Sigma$ binder prefix,
but there is no interaction between those rules and this binder.

\subsection{The $\lambda$-tree approach to syntax encoding}
\label{ssec:ltree}

Those logic programming languages based directly on the sequent
calculus have an elegant and direct technique for specifying
computations in which terms may include bindings.  This technique uses
the three levels of bindings available in sequents: term-level
bindings (in, say, $\lambda$-terms), formula-level bindings ($\forall$
and $\exists$ quantifiers), and proof-level bindings (eigenvariables).
Furthermore, logic specifications are capable of having such bindings
\emph{move} between these different levels.

To illustrate this approach to computing with binders, consider an
encoding of untyped $\lambda$-terms into simply-typed terms.  In
particular, let type $\tm$ be the type of encoded untyped
$\lambda$-terms and let $\app$ and $\abs$ be constants of types
$\tm\ra\tm\ra\tm$ and $(\tm\ra\tm)\ra\tm$, respectively.  The
following three clauses define a function $\lceil\cdot\rceil$ that
translates untyped $\lambda$-terms into terms of type $\tm$.
\[
  \lceil(M N)\rceil = (app~\lceil M\rceil~\lceil N\rceil)
  \qquad
  \lceil{\lambda x}.B\rceil = (abs~(\lambda x. \lceil B\rceil)).
  \qquad
    \lceil x \rceil = x~\hbox{for variable $x$}
\]
Note that bound variables in the untyped $\lambda$-terms correspond to
bound variables in terms of type $\tm$.

Consider now the problem of deciding whether or not an untyped
$\lambda$-term can be given a simple type.  To represent simple types,
we introduce the type $\ty$ and the constant $\arrow:\ty\ra\ty\ra\ty$
that represents the $\ra$ in simple types.
The following logic program specifies the $\typeof{M}{T}$ predicate
that should hold if and only if the untyped $\lambda$-term $M$ has
simple type $T$ (the type of {\sl typeof} is $\tm\ra\ty\ra o$).  Note
that this specification uses a program clause that contains both a
universal quantifier and an implication in its body.
\begin{align*}
\forall B\forall \tvar\forall \tvar' &
  [\typeof{(abs~B)}{(\tvar\ra \tvar')}~
   \backw 
   \forall x(\typeof{x}{\tvar}\supset\typeof{(B~x)}{\tvar'})]
\land\null\\
\forall M\forall N\forall \tvar\forall \tvar' &
   [\typeof{(app~M~N)}{\tvar} \qquad 
   \backw
   \typeof{M}{(\tvar'\ra\tvar)}\land \typeof{N}{\tvar'}]\quad
\end{align*}
Now consider the following combination of inference rules that are
built when type checking a $\lambda$-abstraction.
\[\infer[\hbox{backward chaining}]
  {\Seq{\Sigma:\Delta}
       {\typeof{\lceil{\lambda \underline{x}}.B\rceil}{(\tvar\ra \tvar')}}}
  {\infer[\forall R,\ \supset R]
    {\Seq{\Sigma:\Delta}
         {{\forall \underline{x}}~(\typeof{x}{\tvar}\supset
                      \typeof{\lceil B \rceil}{\tvar'})}}
    {\Seq{\Sigma,\underline{x}:\Delta,\typeof{x}{\tvar}}
         {\typeof{\lceil B \rceil}{\tvar'}}}}
\]
The binding for $x$ moves from the term-level, to the formula-level
(as a quantifier), to the proof-level (as an eigenvariable): these
occurrences are underlined to highlight them.  It is in this sense
that the sequent calculus supports the \emph{mobility of binders}
\cite{miller19jar}: that is, bound variables do not become free,
instead, their scopes move. 

\emph{Higher-order Horn clauses} do not support the movement of
bindings since no universally quantified goals nor eigenvariables are
part of proof search involving them.  In light of this, \lP, which
originally started as an implementation of \hohc was extended to
include both hypothetical and universally quantified goals in order to
support binder mobility.

The term \emph{higher-order abstract syntax} \cite{pfenning88pldi} is
often used to describe systems in which the bindings in data
structures are implemented using bindings in a programming language.
Unfortunately, this term is ambiguous since such identification in the
functional programming setting has almost no relationship with the
approach described above.  For example, if one uses bindings in an
ML-style language, then functions are used to encode the syntax of
terms with bindings.  Such an encoding has many shortcomings
\cite{despeyroux95tlca,hofmann99lics}, and it does not generally
support checking the equality of syntax.  Thus, the approach described
above---providing binder mobility and equality via (at least)
$\alpha$-conversion---has been named the $\lambda$\emph{-tree syntax}
approach to differentiate it from the functional programming
approach~\cite{miller19jar}.

The $\alpha$Prolog system \cite{cheney04iclp} is a logic programming
language with a different approach to encoding and computing
with syntax containing bindings.  Instead of using eigenvariables and
binder mobility, $\alpha$Prolog is based on the logic of Pitts
\citeyear{pitts03ic} which uses the Fraenkel-Mostowski permutation
model of set theory to provide a mechanism for generating and
permuting the names used to encode binders.

\subsection{Unification under a mixed prefix}
\label{ssec:mixed}

Traditional unification can be seen as a technique for proving
formulas of the form  
\[
  \exists x_1\ldots\exists x_n [t_1 = s_1\land\cdots\land t_m=s_m]
  \quad (n\ge 0),
\]
where the quantifier prefix is purely existential.  In principle, such
unification problems are sufficient to consider when building an
interpreter for first-order Horn clauses.  There are at least two ways
in which richer designs of logic programming languages force one to
consider performing unification under a \emph{mixed prefix}, i.e.,
where both existential and universal quantifiers have the conjunction
of equations in their scope.

One such extended design involves replacing first-order terms with
simply-typed $\lambda$-terms, as is the case of \hohc.
The equality theory of such typed terms is generally assumed to
contain not only the $\alpha$, $\beta$, and $\eta$ rules but also 
the $\xi$-rule, which states that the two expressions $\lambda
x.t = \lambda x.s$ and $\forall x. t = s$ are logically equivalent.
Since the forward direction of this equivalence is easily proved, the
force of the $\xi$-rule is the converse.
Using this equivalence, we can show that the unification problem
$\exists y[\lambda z.y =\lambda z.z]$ involving $\lambda$-terms is
equivalent to the mixed prefixed unification problem $\exists y
\forall z [y = z]$, which is an entirely first-order formula (assuming
that the variables $y$ and $z$ have the same primitive type).
Furthermore, these formulas are, in fact, not provable (unless one has
additional axioms stating that the domain of quantification is a
singleton set).
Thus, more generally, the unification of simply-typed $\lambda$-terms
can be seen as having an $\exists\forall$ prefix.

A second extended design arises in the simple use of \fohh.  For
example, let $\Sigma$ be some list of eigenvariables and let $\Pscr$
be the following set of \fohh formulas.
\[
\{ \forall x.~r\;x\;x, \quad 
   \forall y.[(\forall z.~r\;y\;z)\imp q\;y]\}.
\]
It should be clear that there is no proof of $\exists x. q(x)$ from
$\Sigma$ and $\Pscr$.  A proof attempt of this goal can be sketched using
the following arrangement of sequents and pseudo-inference rules.
\[
  \infer[\exists R]{\Seqq{\Sigma}{\Pscr}{\exists x.~q\;x}}
        {\Sigma\vdash X : i &
          \infer[\backchain]{\Seqq{\Sigma}{\Pscr}{q\;X}}{
          \infer[\forall R]{\Seqq{\Sigma}{\Pscr}{\forall z.~r\;X\;z}}{
          \infer[\backchain]{\Seqq{z : i}{\Pscr}{r\;X\;z}}{X = z}}}}
\]
Here, $i$ the a primitive type for quantification.  We use
$X$ as a kind of logic variable: instead of instantiating the
existential quantifier with a term (as is the requirement in Gentzen's
inference rules), we enter $X$ as a kind of hole that we plan to fill
later, but we must remember that whatever fills that hole must be a
term over the variables in $\Sigma$.  Finally, moving upwards through
the series of sequents, we can conclude that we have a proof if
$X$ is instantiated with $z$, which is an eigenvariable that is not a
member of $\Sigma$.  Thus, these two conditions are contradictory, and,
as a result, there is no proof.  Nadathur~\citeyear{nadathur93jar}
describes a unification procedure that works in the presence 
of the quantifier alternations that occur during proof search
with hereditary Harrop formulas.

The general problem of unification of simply-typed $\lambda$-terms
under a mixed prefix can be found in the work by
the author~\citeyear{miller92jsc}, which is 
itself an extension of the earlier work by Huet on unification for
typed $\lambda$-calculus \citeyear{huet75tcs}.  While Skolemization is
often used in automated theorem provers to remove issues
surrounding quantifier alternations, an alternative exists that works
with binder mobility.  It is possible to rotate a universal quantifier
to the right over an existential quantifier: that is, $\forall y
\exists x. B$ and $\exists h \forall y. \subst{(h y)}{x}{B}$ represent the same
unification problem.  In the first, the term $t$ instantiating $x$ can
contain the eigenvariable associated with $y$, while in the second,
$h$ is instantiated with $\lambda y. t$, which, of course, does not
contain $y$ free.  The type of the existentially quantified variable
is raised in this process: in particular, if $y$ has type $\tau$ and
$x$ has type $\tau'$ then $h$ has type $\tau\ra\tau'$.  As a result,
this operation is called \emph{raising}, and it can be used to simplify
all quantifier prenexes to the $\exists\forall$ kind
\cite{miller92jsc}.  (Raising is closely related to the
$\forall$-lifting technique use to deal with eigenvariables in
Isabelle \cite{paulson89jar}.)  To illustrate raising, consider the
following unification problems where $f$ is a function constant of two
arguments.
\begin{align}
  \forall x\exists y\forall z~ [ (f~z~y) & = (f~z~x) ]\nonumber \\
  \forall x\exists y~ [ \lambda z (f~z~y) & = 
                        \lambda z (f~z~x) ] \quad(\xi)\nonumber\\
  \exists h\forall x~ [ \lambda z (f~z~(h~x))& =
                        \lambda z (f~z~x) ]\quad(raising)\nonumber\\
  \exists h~ [\lambda x\lambda z (f~z~(h~x)) & =
              \lambda x\lambda z (f~z~x) ]\quad(\xi)\nonumber
\end{align}
A \emph{solution} is a substitution for the existentially quantified
variables that makes the equated terms the same (modulo
$\alpha\beta\eta$-convertibility).  All of the
above unification problems have their solutions in a one-to-one
correspondence.  In particular, the unique solution for the first
problem is the substitution that maps $y$ to $x$ while the unique
solution for the last problem is the substitution that maps $h$ to
$\lambda x. x$.

In the 1980s and earlier, there were many concerns that higher-order
unification was too complex to allow within the logic programming
setting.  While some concern is justifiable, avoiding all forms of
higher-order unification meant that the full story of unification in
quantificational first-order was not told.  At the same time, early
implementations of higher-order unification in theorem provers
indicated that it was not generally a bottle-neck
\cite{andrews84ams,paulson89jar}.  Part of the reason for the mild
behavior of higher-order unification seems to be that many uses of
higher-order unification tend to belong to the \emph{higher-order
pattern unification} fragment, which, like first-order unification, is
a decidable and unary subset of higher-order unification
\cite{miller91jlc,nipkow93lics}.  In fact, systems such as Twelf
\cite{pfenning99cade}, Teyjus \cite{teyjus.website}, Elpi
\cite{dunchev15lpar}, and Minlog \cite{schwichtenberg06provers} can
encounter arbitrary higher-order unification problems but they only
solve those unification problems that fall within this fragment: in
most practical situations, this approach to higher-order unification
is sufficient.

\subsection{Abstract data types in logic programming}
\label{ssec:adt}

Similar to implications, universal quantifiers in goals can provide
scope for term constructors within goal formulas: exploiting such a
scoping mechanism for constructors provides a logic-based notion of
abstract data type.

Judging from the name ``eigenvariable'', one expects that they
\emph{vary}.  However, eigenvariables do not vary within a cut-free
proof: they act more like constants given a particular scope.  It is
only during cut-elimination that eigenvariables can vary since they
are then substituted by other terms.  Thus, in the setting of proof
search, it makes more sense to view eigenvariables as scoped
constants.

Assume that the variable $y$ is free in the formula $D$ but not in 
$G$.  The interpreter attempting to prove $\forall y(D
\supset G)$ will then introduce a new eigenvariable for $y$, say $k$,
and restrict all the current free variables so that they cannot be
instantiated with terms containing $k$.  The program code $\subst{k}{y}{D}$ can
use the constant $k$ to build data structures.  Of course, if we are
building an interpreter that uses unification, care must be taken to
deal with the fact that some eigenvariables (constants) might be
introduced before or after logic variables are introduced.  We addressed
this issue in Section~\ref{ssec:mixed}.  In the discussions above, the
scope of $y$ is, in a sense, only over $D$ while we needed to use the
universal quantifier $\forall y$ over the compound formula $D\supset
G$, even though $y$ is not free in $G$.  To provide for a more natural
scoping mechanism, note that $(\exists x\; D)\supset G$
and $\forall x(D \supset G)$ are equivalent (in intuitionistic logic)
provided $x$ is not free in $G$. 
Thus, we can use an existential quantifier over program clauses to
limit the scope of constants used in those programs.  Although
$(\exists x\; D)\supset G$ is not a valid hereditary Harrop formula,
it is equivalent to $\forall x(D \supset G)$, which is a valid such
formula.  To allow for the most interesting examples, we shall allow
higher-order quantification for such locally scoped variables.

\begin{figure}
\begin{align*}
\exists emp~\exists stk.~(& &  \exists qu.~( &\\
          & empty\;emp\land\null & 
[\forall l.~& empty\;(qu\;l\;l)] \land\null\\
[\forall s\forall x.~ & enter\;x\;s\;(stk\;x\;s)] \land\null &
[\forall x\forall l\forall k.~& enter\;x\;(qu\;l\;[x|k])\;(qu\;l\;k)]\land\null\\
[\forall s\forall x.~ & remove\;x\;(stk\;x\;s)\;s])&\qquad
[\forall x\forall l\forall k.~& remove\;x\;(qu\;[x|l]\;k)\;(qu\;l\;k)])
\end{align*}
\caption{Two implementations of the predicates $empty$/$enter$/$remove$.}
\label{fig:stack}
\end{figure}

Consider the two existentially quantified conjunctions of Horn clauses
displayed in Figure~\ref{fig:stack}.  In both of those formulas, the
only constants that appear free are the predicates
$empty$, $enter$, and $remove$.  The formula on the left is an
implementation of a stack: here, $emp$ denotes the empty stack, and
$stk$ denotes the non-empty stack constructor.  In this case, the
enter/remove predicates implement the last-in-first-out protocol.  The
formula on the right is an implementation of a queue: here $qu$ forms
a difference list in the usual style familiar to Prolog
programmers~\cite{clocksin94}.  In this case, the enter/remove
predicates implement the first-in-first-out protocol.  Note that by
hiding the internal implementation of the three predicates, it is
possible to change one of these implementations with the other without
the calling code becoming broken.  Of course, the calling code might
well have a different behavior when we swap implementations.

Hiding predicates is also possible using such higher-order
quantification.  For example, the usual way to specify the
relationship between a list and its reverse is often defined using an
auxiliary predicate, which can be hidden using a universal quantifier
in a goal.  Consider the following hereditary Harrop formula.
\begin{align*}
\forall L\forall K.~reverse\;L\;K \backw\null \forall rev.~(&\\
[\forall L.~     & rev\;[]\;L\;L]\imp\null\\
[\forall X\forall L\forall K\forall M.~& rev\;[X|L]\;K\;M \backw
                                        rev\;L\;K\;[X|M]]\imp\null\\
         & \qquad rev\;L\;K\;[])
\end{align*}
To prove an instance of the reverse relationship, this code instructs the
proof search mechanism to create a 
new eigenvariable that plays the role of an auxiliary predicate $rev$
and then loads two Horn clauses that define that auxiliary predicate
before making a call to that auxiliary predicate.  As a result, it is
impossible to access this auxiliary predicate and its code from any
other logic programming clauses that may be in the same context.  More
examples of this approach to abstract data types in logic programming
can be found in~\cite{miller03fcs} and~\cite{miller12proghol}.

\section{Linear logic programming}
\label{sec:llp}

All the previous developments in applying proof theory to logic
programming took place within classical and intuitionistic logic.
When Girard introduced linear logic in \citeyear{girard87tcs}, many
researchers were eager to see if the story behind logic programming
could be extended further using this new logic (see the encyclopedia
article \cite{dicosmo19sep} for an overview of linear logic).
This new logic also seemed to be an extension to both classical and
intuitionistic logic: as a result, there was the promise that linear
logic programming could subsume and extend the various forms of logic
programming we have already described.
Also, the proof theory foundations of the logic programming paradigm had
not provided any hints at how to account for either side-effects or
concurrency: but there were hints that linear logic should 
provide for exactly these missing features. 
Since Girard gave a  simple and clear presentation of linear
logic using the sequent calculus, many researchers started
working on new logic programming designs almost immediately.

Below is a list of several logic programming languages that
incorporate elements of linear logic into their design.  For more
about linear logic programming, the reader is referred to the author's
overview paper \citeyear{miller04llcs}.

\begin{itemize}
\item The LO (linear objects) language designed by Andreoli and
  Pareschi \citeyear{andreoli91ngc} was the first of
  those languages.  LO was a kind of Horn clause logic where atomic
  formulas were generalized to be more like a multiset of
  atomic formulas.  The design provided a natural notion of an
  object-as-process that has a built-in notion of inheritance.

\item Lolli is a simple extension to hereditary Harrop formulas
  \cite{hodas91lics,hodas94ic}.  Essentially, the linear implication
  $\limp$ is allowed to appear in the same way as the intuitionistic
  implication can appear: at the top-level of both definite clauses
  and goals.  Lolli had the property that if a program never uses
  $\limp$ as a goal formula, then proofs and proof search are
  essentially the same as when using intuitionistic logic.  A new
  feature that Lolli provides over \lP is a mechanism for describing
  state and state change, including database updates and retraction.

\item The Lygon system of Harland and Pym \citeyear{harland96amast} was
  designed following a proof-theoretic analysis of goal-directed proof
  in linear logic \cite{pym94jlp}.  The application areas of Lygon and
  Lolli overlap significantly.

\item The language ACL by Kobayashi and Yonezawa
  \cite{kobayashi93ilps,kobayashi94fac} captures simple notions of
  asynchronous communication by identifying the send and read
  primitives with two complementary linear logic connectives.

\item Lincoln and Saraswat developed a linear logic version of
  concurrent constraint programming \cite{lincoln93tr,saraswat93tr},
  and Fages, Ruet, and Soliman have analyzed similar extensions to
  the concurrent constraint paradigm \cite{ruet97csl,fages98lics}.

\item The Forum language \cite{miller96tcs,bruscoli06tcs} is
  essentially a presentation of linear logic that allows for all of
  linear logic to be considered as an abstract logic programming
  language.  The proof-theoretic analysis of Forum required lifting
  the notion of goal-directed proofs to deal with multiple-conclusion
  sequents.  Forum can be seen as the result of merging LO and Lolli.
\end{itemize}

An early observation about linear logic is that it supports 
multiset rewriting in a rather direct fashion.  Thus, linear logic
programming can encode both Petri nets
\cite{gunter89upenn,kanovich95apal} and the process calculi, such as
the $\pi$-calculus \cite{miller92welp}.

To illustrate how sequent calculus can be used to encode a small
fragment of linear logic (the fragment that deals with $\supset$,
$\limp$, and $\forall$), we present the \LL proof system in
Figure~\ref{fig:ll}.  We continue to use $\supset$ to denote
(intuitionistic) implication and introduce Girard's linear implication
$\limp$.  Part of the informal meaning of linear implication is that a
proof of $B\limp C$ is a proof of $C$ in which the assumption $B$ is
used exactly once.  The corresponding informal meaning of the
intuitionistic implication is that a proof of $B\imp C$ is a proof of
$C$ in which the assumption $B$ is used any number of times, including
zero.  To permit these two different accounting methods for
assumptions, the left-hand side of sequents is divided into two zones.
In the sequent $\lSeq{\Gamma}{\Delta}{E}$, the context $\Gamma$ holds
the assumptions under the unbounded-use accounting, and the context
$\Delta$ contains the assumptions under the single-use accounting: we
refer to $\Gamma$ as the \emph{unbounded zone} and $\Delta$ as the
\emph{bounded zone}. 
Hodas and the author \citeyear{hodas94ic} proved that \LL (over the
same connectives) is sound and (relatively) complete for linear logic.

Girard's original presentation of linear logic \citeyear{girard87tcs}
did not rely on using the two implications $\limp$ and $\imp$.
Instead, the implication $B\imp C$ was defined as $\bang B\limp C$,
where $\bang$ is one of linear logic's \emph{exponentials}.  A
formula marked by a $\bang$ can be contracted and weakened when it
appears on the left side of a sequent arrow.  Dually, a formula marked
by the other exponential $\quest$ can be contracted and weakened when
it appears on the right.  With these exponentials, linear logic can
encode both classical and intuitionistic logics.  We have chosen not
to use the exponentials of linear logic here, but if we did introduce
it, then the sequent $\lSeq{B_1,\ldots,B_n}{\Delta}{E}$ could be
rewritten as $\Seq{\bang B_1,\ldots,\bang B_n,\Delta}{E}$.

\begin{figure}
\begin{flushleft}
  \textsc{Structural rules}
\end{flushleft}
\[
  \infer[\contrL]{\lSeq{\Gamma,B}{\Delta}{E}}{\lSeq{\Gamma,B,B}{\Delta}{E}}
  \qquad
  \infer[\weakL]{\lSeq{\Gamma,B}{\Delta}{E}}{\lSeq{\Gamma}{\Delta}{E}}
  \qquad
  \infer[\dereliction]{\lSeq{\Gamma,B}{\Delta}{E}}{\lSeq{\Gamma}{B,\Delta}{E}}
\]
\begin{flushleft}
  \textsc{Identity rules}
\end{flushleft}
\[
  \infer[\init]{\lSeq{\cdot}{B}{B}}{}
  \qquad\qquad
  \infer[\cut]{\lSeq{\Gamma_1,\Gamma_2}{\Delta_1,\Delta_2}{E}}
              {\lSeq{\Gamma_1}{\Delta_1}{B}\qquad
               \lSeq{\Gamma_2}{\Delta_2,B}{E}}
\]
\begin{flushleft}
  \textsc{Introduction rules}
\end{flushleft}
\[
  \infer[\impL]
        {\lSeq{\Gamma_1,\Gamma_2}{\Delta,B_1 \supset B_2}{E}}
        {\lSeq{\Gamma_1}{\cdot}{B_1}\qquad \lSeq{\Gamma_2}{\Delta,B_2}{E}}
  \qquad\qquad
  \infer[\impR]{\lSeq{\Gamma}{\Delta}{B_1 \imp B_2}}{\lSeq{\Gamma,B_1}{\Delta}{B_2}}
\]
\[
  \infer[\limpL]
        {\lSeq{\Gamma_1,\Gamma_2}{\Delta_1,\Delta_2,B_1 \limp B_2}{E}}
        {\lSeq{\Gamma_1}{\Delta_1}{B_1}\qquad \lSeq{\Gamma_2}{\Delta_2,B_2}{E}}
  \qquad\qquad
  \infer[\limpR]{\lSeq{\Gamma}{\Delta}{B_1 \limp B_2}}{\lSeq{\Gamma}{\Delta,B_1}{B_2}}
\]
\[
  \infer[\forallL]{\lSeq{\Gamma}{\Delta,\forall x.B}{E}}{\lSeq{\Gamma}{\Delta,\subst{t}{x}{B}}{E}}
  \qquad\qquad
  \infer[\forallR]
        {\lSeq{\Gamma}{\Delta}{\forall x.B}}{\lSeq{\Gamma}{\Delta}{\subst{y}{x}{B}}}
\]
\caption{The \LL proof system for $\supset$, $\limp$, and $\forall$.
         In the $\forallR$ rule, the variable $y$ is not free in the
         conclusion of that rule.}
\label{fig:ll}
\end{figure}

When comparing the subset of the \LJ proof system in
Figure~\ref{fig:lj} with the \LL proof system in Figure~\ref{fig:ll},
we see that the contraction and weakening rules are available only in
the unbounded zone, that the $\impR$ rule adds its hypothesis to the
unbounded zone, and that the $\limpR$ rule adds its hypothesis to the
bounded zone.  Finally, also note that the two left-introduction rules
for implication treat their unbounded zones \emph{multiplicatively},
meaning that every side-formula occurrence in the bounded context of
the conclusion occurs in a bounded zone of exactly one premise.
Furthermore, in the $\impL$ rule, the bounded zone of the left premise
must be empty.  Also, note that the only formula occurrences that are
introduced on the left occur in the bounded zone.  The \dereliction
rule is responsible for moving a formula in the unbounded zone to the
bounded zone.

Figure~\ref{fig:llprime} contains the simplification \LLp of \LL in
which we remove the cut-rule (since we are generally interested here
in cut-free proofs) and in which we fold the weakening and contraction
rules into other rules so these rules are never explicitly invoked.
They are still present in this simplified proof system, however.  In
particular, the $\init$ rule allows the unbounded zone to be non-empty
(since weakenings can be used to empty that zone) and the two
implication-left rules keep the the unbounded zone the same in the
premises and the conclusion.  Also, the $\absorb$ rule links
contraction with the $\dereliction$ rule.

\begin{figure}
\[
  \infer[\init]{\lSeq{\Gamma}{B}{B}}{}
  \qquad
  \infer[\absorb]{\lSeq{\Gamma,B}{\Delta}{E}}{\lSeq{\Gamma,B}{B,\Delta}{E}}
\]
\[
  \infer[\impL]
        {\lSeq{\Gamma}{\Delta,B_1 \supset B_2}{E}}
        {\lSeq{\Gamma}{\cdot}{B_1}\qquad \lSeq{\Gamma}{\Delta,B_2}{E}}
  \qquad\qquad
  \infer[\impR]{\lSeq{\Gamma}{\Delta}{B_1 \imp B_2}}{\lSeq{\Gamma,B_1}{\Delta}{B_2}}
\]
\[
  \infer[\limpL]
        {\lSeq{\Gamma}{\Delta_1,\Delta_2,B_1 \limp B_2}{E}}
        {\lSeq{\Gamma}{\Delta_1}{B_1}\qquad \lSeq{\Gamma}{\Delta_2,B_2}{E}}
  \qquad\qquad
  \infer[\limpR]{\lSeq{\Gamma}{\Delta}{B_1 \limp B_2}}{\lSeq{\Gamma}{\Delta,B_1}{B_2}}
\]
\[
  \infer[\forallL]{\lSeq{\Gamma}{\Delta,\forall x.B}{E}}{\lSeq{\Gamma}{\Delta,\subst{t}{x}{B}}{E}}
  \qquad\qquad
  \infer[\forallR]
        {\lSeq{\Gamma}{\Delta}{\forall x.B}}{\lSeq{\Gamma}{\Delta}{\subst{y}{x}{B}}}
\]
\caption{The \LLp proof system results from building into the \LL
  proof system the structural rules of \weakL and \contrL.}
\label{fig:llprime}
\end{figure}

To illustrate how linear logic provides for new logic programs with
new dynamics, consider the following two linear logic formulas.
\begin{align*}
  \forall G. (\swon  \limp G) \limp (\swoff \limp toggle\;G)\\
  \forall G. (\swoff \limp G) \limp (\swon  \limp toggle\;G)
\end{align*}
Linear logic contains the conjunction $\otimes$ (pronounced
``tensor'') for which the equivalence $(A\limp B\limp
C)\equiv((A\otimes B)\limp C)$ holds.  Following
the Lolli language conventions~\cite{hodas94phd,hodas94ic},
we write the $\otimes$ as a comma and the converse of $\limp$ as 
\texttt{:-}.  As a result, these two formulas can be written in the
following Prolog-like style. 
\begin{verbatim}
toggle(G) :- sw off, (sw on  -o G).
toggle(G) :- sw on,  (sw off -o G).
\end{verbatim}
Using the proof system \LLp (Figure~\ref{fig:llprime}), we can build
the following partial proof.
\[
  \infer[\absorb]{\lSeq{\Gamma}{\swon,\Delta}{\toggle{g}}}{
  \infer[\forallL]{\lSeq{\Gamma}
                        {\forall G.(\swoff\limp G)\limp(\swon\limp\toggle{g}),\swon,\Delta}
                        {\toggle{g}}}{
  \infer[\limpL\times 2]{\lSeq{\Gamma}{(\swoff\limp g)\limp(\swon\limp\toggle{g}),\swon,\Delta}{\toggle{g}}}
        {\infer[\limpR]{\lSeq{\Gamma}{\Delta}{\swoff\limp g}}{\lSeq{\Gamma}{\swoff,\Delta}{g}} &
         \infer[\init]{\lSeq{\Gamma}{\swon}{\swon}}{} &
         \infer[\init]{\lSeq{\Gamma}{\toggle{g}}{\toggle{g}}}{}}}}
\]
This derivation (and the analogous one using the other formula for
\texttt{toggle}) essentially interprets these two clauses for
\texttt{toggle} as the following two admissible rules.
\[
\infer{\lSeq{\Gamma}{\swon,\Delta}{\toggle{g}}}{\lSeq{\Gamma}{\swoff,\Delta}{g}}
\hbox{\qquad and \qquad}
\infer{\lSeq{\Gamma}{\swoff,\Delta}{\toggle{g}}}{\lSeq{\Gamma}{\swon,\Delta}{g}}
\]
Thus, the process of reducing the goal $\toggle{g}$ to $g$ will flip
the switch's value stored as the argument of (the presumably unique)
\textsl{sw}-atom and will affect no other formula in the bounded or
unbounded zones.  We can attempt something similar using
intuitionistic logic and hereditary Harrop: for example, consider
the following specification for toggle.
\begin{align*}
  \forall G. (\swon  \imp G) \imp (\swoff \imp toggle\;G)\\
  \forall G. (\swoff \imp G) \imp (\swon  \imp toggle\;G)
\end{align*}
A proof fragment in intuitionistic logic starting with one of these
formulas will look as follows.
\[
  \infer[\absorb]{\Seq{\Gamma,\swon}{\toggle{g}}}{
  \infer[\forallL]{\Seq{\Gamma,\forall G.(\swoff\imp G)\imp(\swon\imp\toggle{g}),\swon}
                        {\toggle{g}}}{
  \infer[\hbox{several }\contrL]{\Seq{\Gamma,(\swoff\imp g)\imp(\swon\imp\toggle{g}),\swon}{\toggle{g}}}{
  \infer[\impL\times 2]{\Seq{\Gamma,\Gamma,(\swoff\imp g)\imp(\swon\imp\toggle{g}),\swon,\swon}{\toggle{g}}}
        {\infer[\impR]{\Seq{\Gamma,\swon}{\swoff\imp g}}{\Seq{\Gamma,\swon,\swoff}{g}} &
         \infer[\init]{\Seq{\Gamma,\swon}{\swon}}{} &
         \infer[\init]{\Seq{\toggle{g}}{\toggle{g}}}{}}}}}
\]
In this setting, we get the admissible rules
\[
\vcenter{\infer{\Seq{\Gamma,\swon}{\toggle{g}}}{\Seq{\Gamma,\swon,\swoff}{g}}}
\hbox{\qquad and \qquad}
\vcenter{\infer[,]{\Seq{\Gamma,\swoff}{\toggle{g}}}{\Seq{\Gamma,\swon,\swoff}{g}}}
\]
which is not what we expect from a proper switch.

The mechanism behind this simple example can easily be expanded to
perform multiset rewriting.  Let $H$ be the \emph{multiset rewriting
  system} $\{\tup{L_i,R_i}\sep i\in I\}$ where for each $i\in I$ (a
finite index set), $L_i$ and $R_i$ are finite multisets. Define the
relation $M\Lra_H N$ on finite multisets to hold if there is some
$i\in I$ and some multiset $C$ such that $M$ is $C\uplus L_i$ and $N$
is $C\uplus R_i$.  Let $\Lra_H^*$ be the reflexive and transitive
closure of $\Lra_H$.

The $H$ rewrite system can be encoded as a multiset of linear logic
formulas as follows: If $H$ contains the pair
$\tup{\{a_1,\ldots,a_n\},\{ b_1,\ldots,b_m\}}$ then this pair is
encoded as the clause
\begin{verbatim}
loop :- item a1,   ...,   item an,
        (item b1 -o ... -o item bm -o loop).
\end{verbatim}
If either $n$ or $m$ is zero, the appropriate portion of the formula is 
deleted.   Here \texttt{item} is a predicate of one argument that is
used to inject multiset items into atomic formulas.
Operationally, this clause (destructively) reads the $a_i$'s out of the 
bounded context, loads the $b_i$'s into that context, and then attempts another rewrite.
Let $\Gamma_H$
be the set resulting from encoding each pair in $H$.
For example, if $H=\{\tup{\{a,b\},\{b,c\}},\tup{\{a,a\},\{a\}}\}$ then $\Gamma_H$ is
the set of clauses
\begin{verbatim}
loop  :-   item a, item b, (item b -o item c -o loop).
loop  :-   item a, item a, (item a -o loop).
\end{verbatim}
The following holds about this encoding of
multiset rewriting: the relation
\[\{a_1,\ldots,a_n\}\Lra_H^* \{b_1,\ldots,b_m\}\]
holds if and only if sequent
$\lSeq{\Gamma}{\Delta,\itm{a_1},\ldots,\itm{a_n}}{\hbox{\textsl{loop}}}$
can be derived from the sequent
$\lSeq{\Gamma}{\Delta,\itm{b_1},\ldots,\itm{b_m}}{\hbox{\textsl{loop}}}$.

As these examples illustrate, the existence of formulas with limited
use increases the expressiveness of linear logic programs.  Along
with that increase in expressiveness comes an increase in the cost of
doing proof search.  In particular, consider the $\limpL$ inference
rule from Figure~\ref{fig:llprime}, namely,
\[
  \infer[\limpL.]
        {\lSeq{\Gamma}{\Delta_1,\Delta_2,B_1 \limp B_2}{E}}
        {\lSeq{\Gamma}{\Delta_1}{B_1}\qquad \lSeq{\Gamma}{\Delta_2,B_2}{E}}
\]
When reading this inference from bottom to top, one must decide to
take the side-formulas to $B_1\limp B_2$ within the bounded context,
say $\Delta$, and split that multiset into $\Delta_1$ (the formulas
given to the left-premise) and $\Delta_2$ (the formulas given to the
right-premise).  If $\Delta$ contains $n$ occurrences of distinct
formulas, then there are $2^n$ possible splittings
available.  It is possible to implement interpreters for linear logic
programming languages in which $\Delta$ is not split immediately into
$\Delta_1$ and $\Delta_2$ but rather all of $\Delta$ is given to, say,
the proof attempt on the left premise.  If that proof attempt is
successful, then $\Delta_1$ is taken to be those formulas consumed
from the bounded context during that attempt.  The formulas that
result from removing $\Delta_1$ from $\Delta$ yield $\Delta_2$.  Such
a lazy splitting approach has been used in various implementations of
linear logic programming languages \cite{hodas94ic,cervesato00tcs}.

As we have seen, much of the novel expressiveness of linear logic
programmings comes from their ability to express multiset rewriting.
In the specifications we have presented above, there is, however, only
one multiset that is subjected to rewriting, and that is the multiset
that forms the zone $\Delta$ in sequents of the form
$\lSeq{\Gamma}{\Delta}{E}$.  An even more expressive framework would
allow the left-hand side to be divided into multiple zones
representing multiple multisets.  In that setting, different parts of
a logic program could use different multisets for different purposes.
It turns out that just such multiple-zone sequents are possible in
linear logic by noting that the exponentials of linear logic are not
\emph{canonical} logical connectives.  To explain what we mean by
\emph{canonical}, consider adding to linear logic the logical
connective $\with'$ which has the same inference rules that exist
for $\with$.  In such an extended logic and proof system, it is easy
to prove that $B\with C$ and $B\with' C$ are logically equivalent
formulas.  As a result, we say that $\with$ is a \emph{canonical
logical connective}.  Nothing is gained by adding such a variant of
$\with$.

It is easy to show that all the connectives of linear logic are
canonical except for the exponentials $\bang$ and $\quest$.  That is,
if we add a blue $\nbang{b}$ and a red $\nbang{r}$ to linear logic and
give them each the same inference rules that exist for $\bang$, we
then have a more expressive logic.  Furthermore, it is possible to
allow explicitly the contraction and weakening rules to be applicable
for formulas explicitly marked by, say, $\nbang{b}$ but not for
$\nbang{r}$.  Danos \etal\ \citeyear{danos93kgc} proposed a linear
logic system with such non-canonical exponentials and illustrated
their uses in the framework of the Curry-Howard correspondence.  These
non-canonical exponentials are now called \emph{subexponentials}
\cite{nigam09ppdp}.  As we have seen, the difference between the two
zones on the left of sequents in $\lSeq{\Gamma}{\Delta}{E}$ comes down
to the fact that the formulas in $\Gamma$ should be considered as
having $\bang$ attached to them while the formulas in $\Delta$ do not
have $\bang$ attached.  Thus, the existence of $n$ different
subexponentials can now encode $n+1$ zones on the left-hand side of
sequents, and some of these zones will allow weakening and contraction
(such as the $\Gamma$ zone), and other zones will allow neither of
these structural rules (such as the $\Delta$ zone).  Similarly,
expressions of the form $\nbang{b}B \limp C$ and $\nbang{r}B \limp C$
would provide new kinds of implications.  The additional
expressiveness of subexponentials in the logic programming setting has
been developed a great deal in recent years: see the papers
\cite{nigam09phd,nigam09ppdp,chaudhuri10csl,nigam10lsfa,olarte15tcs,despeyroux16lsfa,kanovich19mscs}.
Subexponentials have also been used to encode concurrent process
calculi \cite{nigam17tcs} and aspects of Milner's bigraphs
\cite{chaudhuri15lpar}.

\section{Focusing and polarities}
\label{sec:focusing}

\subsection{Extending two phases to linear logic}

Once Girard introduced linear logic in \citeyear{girard87tcs},
Andreoli generalized the two-phase structure of uniform
proofs (see Section~\ref{sec:gdp}) with the design of a \emph{focused
proof system} for linear logic~\cite{andreoli90phd,andreoli92jlc}.
Two important
insights distinguish focused proofs from uniform proofs.  First,
Andreoli's original focused proof system was defined for linear logic,
which is more expressive than intuitionistic logic and contains an
involutive negation.  Second, and more importantly, Andreoli's phases
were based on the notion of \emph{invertibility} and
\emph{non-invertibility}.  From the proof search point-of-view,
invertible rules can be applied in a
\emph{don't-care-nondeterministic} fashion, whereas the non-invertible
rules can be applied in a \emph{don't-know-nondeterministic} fashion.
As it turns out, the distinction between invertible and non-invertible
inference rules is more fundamental than the distinction between
left-hand side and right-hand side, especially in linear logic where
the systematic use of negation means that all sequents can be assumed
to be one-sided.

\begin{figure}
\[
  \infer[\init]{\fSeqq{\Sigma}{\Gamma}{\cdot}{A}{A}}{}
  \qquad
  \infer[\decide]{\lSeqq{\Sigma}{\Gamma}{\Delta,B}{A}}
                 {\fSeqq{\Sigma}{\Gamma}{\Delta}{B}{A}}
  \qquad
  \infer[\decideb]{\lSeqq{\Sigma}{\Gamma,B}{\Delta}{A}}
                  {\fSeqq{\Sigma}{\Gamma,B}{\Delta}{B}{A}}
\]
\[
  \infer[\impL]
        {\fSeqq{\Sigma}{\Gamma}{\Delta}{B_1 \supset B_2}{A}}
        {\lSeqq{\Sigma}{\Gamma}{\cdot}{B_1}\qquad
         \fSeqq{\Sigma}{\Gamma}{\Delta}{B_2}{A}}
  \qquad\qquad
  \infer[\impR]{\lSeqq{\Sigma}{\Gamma}{\Delta}{B_1 \imp B_2}}
               {\lSeqq{\Sigma}{\Gamma,B_1}{\Delta}{B_2}}
\]
\[
  \infer[\limpL]
        {\fSeqq{\Sigma}{\Gamma}{\Delta_1,\Delta_2}{B_1 \limp B_2}{A}}
        {\lSeqq{\Sigma}{\Gamma}{\Delta_1}{B_1}\qquad
         \fSeqq{\Sigma}{\Gamma}{\Delta_2}{B_2}{A}}
  \qquad\qquad
  \infer[\limpR]{\lSeqq{\Sigma}{\Gamma}{\Delta}{B_1 \limp B_2}}
                {\lSeqq{\Sigma}{\Gamma}{\Delta,B_1}{B_2}}
\]
\[
  \infer[\forallL]{\fSeqq{\Sigma}{\Gamma}{\Delta}{\forall_\tau x.B}{A}}
                  {\Sigma\vdash t:\tau\qquad
                   \fSeqq{\Sigma}{\Gamma}{\Delta}{\subst{t}{x}{B}}{A}}
  \qquad\qquad
  \infer[\forallR]
        {\lSeqq{\Sigma}{\Gamma}{\Delta}{\forall_\tau x.B}}
        {\lSeqq{\Sigma,y:\tau}{\Gamma}{\Delta}{\subst{y}{x}{B}}}
\]
\caption{The \LLf proof system.  In these sequents, $A$ denotes a
  syntactic variable for atomic formulas.}
\label{fig:llfocus}
\end{figure}

Figure~\ref{fig:llfocus} presents a focused version of the \LLp proof
system of Figure~\ref{fig:llprime}.  This proof system, which is a
subset of the $\Fscr$ proof system in~\cite{miller96tcs}, contains two
kinds of sequents.  Sequents of the form $\lSeqq{\Sigma}{\Gamma}{\Delta}{B}$
are essentially the sequents that appear in \LLp but in the \LLf
sequents of this style can only be the conclusion of
right-introduction rules or the $\decide$ or $\decideb$ rules.  The second
kind of sequent is of the 
form $$\fSeqq{\Sigma}{\Gamma}{\Delta}{B}{A}$$ where $A$ is an atomic formula.
Here, the $\DOWNarrow$ provides the left-hand side of a sequent with
an additional zone between $\DOWNarrow$ and $\vdash$: this
new zone always contains exactly one formula.  Sequents containing a
$\DOWNarrow$ are called \emph{focused sequents} and they can only be
the conclusion of left-introduction rules or the $\init$ rule.  Thus,
we can see two phases in focused proof construction.  One phase
involves only sequents containing $\DOWNarrow$ and having an atomic
right-hand side.  The other phase involves only sequents that do not
contain $\DOWNarrow$.  If we revisit the derivation in
Section~\ref{sec:gdp} containing underlined formulas, it is easy to
rewrite that derivation in the \LLf proof system in such a way that
the underlined formulas correspond to the formulas next to the
$\DOWNarrow$.  The $\DOWNarrow$-phase corresponds to backward
chaining and the phase without the $\DOWNarrow$ corresponds to the
goal-reduction phase of uniform proofs.

Andreoli's focused proof system was for a version of linear logic that
did not include the $\limp$ and $\imp$ implications and, as a result,
that proof system was one-sided (all formulas are placed on the right of the
sequent arrow).  However, it is possible to revise Andreoli's proof
system to include both implications and identify the
$\DOWNarrow$-phase with left-introduction rules and the
$\DOWNarrow$-free phase with right-introduction rules.  This
reorganization of focused linear logic proofs is called the Forum
logic programming language \cite{miller94lics,miller96tcs}.  This
presentation of linear logic allows one to view logic programming
using Horn clauses (Section~\ref{sec:gdp}), hereditary Harrop formulas
(Section~\ref{sec:modules}), and Lolli (Section~\ref{sec:llp}) all as
subsets of just the one, large logic programming language.
The Forum presentation of linear logic allows us to conclude
that all of linear logic is an abstract logic programming
language~\cite{miller96tcs}. 

\subsection{The dynamics of an abstract logic programming language}
\label{ssec:dynamics}

Recall that sequents are used to capture the state of a logic
programming computation: that is, the sequent $\Seqq\Sigma{\Pscr}{G}$
represents a configuration where the current logic program is $\Pscr$,
the current goal is $G$, and the current signature of eigenvariables
(scoped constants) is $\Sigma$.  A natural and high-level
characterization of logic programming languages is captured by the question: How
richly can these configurations change during the search for a proof?
In other words, if $\Seqq{\Sigma}{\Pscr}{G}$ is the root of a
derivation and if $\Seqq{\Sigma'}{\Pscr'}{G'}$ is a sequent occurring
above the root in that derivation, what is the relationship between
$\Sigma$ and $\Sigma'$, between $\Pscr$ and $\Pscr'$, and between $G$
and $G'$?
Focused proof systems, such as the \LLf proof system of
Figure~\ref{fig:llfocus}, provides a natural and simple way to answer
this question.  In particular, call a sequent of the form
$\lSeqq{\Sigma}{\Gamma}{\Delta}{A}$, for atomic $A$, a \emph{border
  sequent}.  Such sequents occur between the goal reduction phase and
the backward-chaining phase.  We can limit our questions about
dynamics to just such border sequents:

\par\smallskip\hangindent 20pt\quad
if $\lSeqq{\Sigma}{\Gamma}{\Delta}{A}$ is the root of a derivation and
if $\lSeqq{\Sigma'}{\Gamma'}{\Delta'}{A'}$
is a border sequent occurring above
the root in that derivation, what is the relations between $\Sigma$
and $\Sigma'$, between $\Gamma$ and $\Gamma'$, between $\Delta$ and
$\Delta'$, and between $A$ and $A'$?
\par\smallskip

If $\Gamma$ is a multiset of Horn clauses, then we can
immediately say that $\Sigma=\Sigma'$, $\Gamma=\Gamma'$,
$\Delta=\Delta'$ and $\Delta$ must be, in fact, the empty multiset.
Only the relationship between $A$ and $A'$ can be rich.  That is, the
left-hand side of the sequent is constant and global during the entire
computation.  The only dynamics of computation must take place within
atomic formulas.

If $\Gamma$ is a multiset of hereditary Harrop formulas,
then we can immediately say that $\Sigma\subseteq\Sigma'$,
$\Gamma\subseteq\Gamma'$, $\Delta=\Delta'$ and $\Delta$ must be, in
fact, the empty multiset.  Thus, slightly richer dynamics can take
place in this setting since both the signature and the logic program
can grow as proof search progresses.

If $\Gamma$ is a multiset of any formulas using
$\forall$, $\imp$, and $\limp$, then we can immediately say again that
$\Sigma\subseteq\Sigma'$ and $\Gamma\subseteq\Gamma'$ hold but that
there is no simple relationship between $\Delta$ and $\Delta'$.  The
relationship between these two multiset sets can be, essentially,
arbitrary and depends on the nature of the logic programs available in
that sequent.

\subsection{Polarization applied to classical and intuitionistic logics}

A standard presentation of linear logic does not involve implications
(neither $\imp$ nor $\limp$) and, as such, a sequent calculus for it
can use one-sided sequents.  In that setting, it turns out that the
right-introduction rules for a given connective are invertible if and
only if the right-introduction rules for the De Morgan dual of that
connective are non-invertible.  Such a property suggests introducing
the notion of the \emph{polarities} of a logic connective
\cite{girard91mscs,andreoli92jlc}.  In particular, a logical
connective is \emph{negative} if its right-introduction rule is
invertible, and a logical connective is \emph{positive} if it is the
De Morgan dual of a negative connective.

Once polarity and focusing are described in terms of invertibility of
inference rules, it is possible to apply them to proof systems
in other logics.  For example, Girard~\citeyear{girard91mscs}, Danos
et al.~\citeyear{danos93wll}, Curien and
Munch-Maccagnoni~\citeyear{curien10tcs}, and
Wadler~\citeyear{wadler03icfp} applied these concepts to classical
logic in order to develop well-structured notions of functional
programming in classical logic (via the Curry-Howard correspondence).

These concepts have also been applied in intuitionistic logic.  For
example, Herbelin~\citeyear{herbelin95phd} and Dyckhoff and
Lengrand~\citeyear{dyckhoff06cie} developed focused proof systems for
intuitionistic logic while the Ph.D. theses of Howe
\citeyear{howe98phd} and Chaudhuri \citeyear{chaudhuri06phd} explored
more variations on focused proof systems for both linear and
intuitionistic logics.  Liang and the author have developed focused
proof systems for classical and intuitionistic first-order
logics---called \LKF and \LJF, respectively---which can account for
these various, earlier focused proof systems~\cite{liang07csl,liang09tcs}.

\begin{figure}
\[
  \infer{\Seq{\Gamma}{B_1 \supset B_2}}{\Seq{\Gamma,B_1}{B_2}}
  \qquad\qquad
  \infer[\hbox{$y$ not free in conclusion}]
        {\Seq{\Gamma}{\forall x.B}}{\Seq{\Gamma}{\subst{y}{x}{B}}}
\]
\[
  \infer{\jLf{\Gamma}{\forall x.B}{A}}{\jLf{\Gamma}{\subst{t}{x}{B}}{A}}
  \qquad
  \infer{\jLf{\Gamma}{B_1 \supset B_2}{A}}
        {\jRf{\Gamma}{B_1}\qquad \jLf{\Gamma}{B_2}{A}}
\]
\[
 \labelleft{Decide:}\qquad
 \infer[\kern -2pt D_l]{\Seq{\Gamma,N}{A}}{\jLf{\Gamma,N}{N}{A}}
 \qquad
 \infer[\kern -2pt D_r]{\Seq{\Gamma}{P}}{\jRf{\Gamma}{P}}
\]
\[
 \labelleft{Release:}\qquad
  \infer[R_l]{\jLf{\Gamma}{P}{A}}{\Seq{\Gamma,P}{A}}
  \qquad
  \infer[R_r]{\jRf{\Gamma}{N}}{ \Seq{\Gamma}{N}}
\]
\[
 \labelleft{Initial:}\qquad
  \infer[I_l]{\jLf{\Gamma}{N}{N}}{N \; \mbox{atomic}}
  \qquad
  \infer[I_r]{\jRf{\Gamma,P}{P}}{P \; \mbox{atomic}}
\]
Here,
$A$ is an atomic formula (of either polarity), 
$P$ is a positive (atomic) formula, and 
$N$ is a negative formula.

\caption{The \LJFp proof system: \LJF restricted to only $\supset$ and
  $\forall$.}
\label{fig:ljfp}
\end{figure}

To illustrate the use of \LJF in our setting, consider the \LJFp proof
system given in Figure~\ref{fig:ljfp}.  This proof system is a subset
of \LJF and resembles the \LLf.  
There are three kinds of sequents.
\begin{enumerate}
\item unfocused sequents: $\Seq{\Gamma}{B}$ 
\item left focused sequents: $\jLf{\Gamma}{B}{A}$,  with focus $B$
\item right focused sequents: $\jRf{\Gamma}{A}$,  with focus $A$
\end{enumerate}
Replacing $\DOWNarrow$ on the left with a comma and dropping
$\DOWNarrow$ on the right yields a regular
sequent.

The formulas of \LJFp are given polarity as follows.  Since the right
rules for $\imp$ and $\forall$ are invertible, formulas of the form
$B_1\supset B_2$ and $\forall x. B$ are \emph{negative}.  We shall
assign a polarity also to atomic formulas by allowing them to have an
arbitrary (but fixed) polarity.  Thus, atomic formulas can be either
positive or negative.  In the more general setting, \LJF has more
positive formulas (including disjunctions and existential
quantifiers), but in this setting where we only consider implications
and universal quantifiers, only atoms can be positive formulas.

Uniform proofs, when restricted to the logical connectives for
implication and universal quantification, correspond to focused proofs
where all atomic formulas are polarized negatively.  

The following result about \LJFp follows from the more general
results for \LJF given by Liang and the author~\citeyear{liang09tcs}.
Let $B$ be a first-order formula built from
atomic formulas, $\forall$, and $\supset$.
\begin{itemize}
  \item If $\Gamma\vdash B$ is provable in \LJ then for every
    polarization of atomic formulas, the sequent $\Seq{\Gamma}{B}$ is
    provable in \LJFp.
  \item If atoms are given some polarization and $\Seq{\Gamma}{B}$ is
    provable in \LJFp, then $\Gamma\vdash B$ is provable in \LJ. 
\end{itemize}
An immediate conclusion of this result is that the choice of the
polarity of atoms does not affect provability.  As we shall see next,
that choice can have a big impact on the structure of proofs.

\subsection{Characterizing forward and backward chaining}
\label{ssec:fcbc}

In the Curry-Howard correspondence, different control regimes for
evaluation (e.g., call-by-value and call-by-name) can be explained by
different choices in polarizations in intuitionistic logic formulas
\cite{brocknannestad15ppdp,espiritosanto16lsfa}.  In the proof search
setting, two familiar control strategies---top-down and
bottom-up---can similarly be explained by using two different
polarizations of atomic formulas with the \LJF proof system.  For
example, consider the following partial derivation within \LJFp.
\[
\infer[\forall L\times 3]
      {\Gamma\mathrel\DOWNarrow\forall x\forall y\forall z(\R x y\supset \R y z\supset \R x z)\vdash A}{
\infer[\supset L]{\Gamma\mathrel\DOWNarrow \R a b\supset \R b c\supset \R a c\vdash A}{
   \deduce{\Gamma\vdash \R a b\mathrel\DOWNarrow}{\strut\Xi_1} & 
   \infer[\supset L]{\Gamma\mathrel\DOWNarrow \R b c\supset \R a c\vdash A}{
      \deduce{\Gamma\vdash \R b c\mathrel\DOWNarrow}{\strut\Xi_2} &
      \deduce{\Gamma\mathrel\DOWNarrow \R a c\vdash A}{\strut\Xi_3}}}}
\]
Here, $A$ is some atomic formula, $a,b,c$ are three terms, and the
formula under focus in the concluding sequent states that the $r$
relation is transitive.   To complete the construction of this
focused proof, we need to know the polarity of the atomic formulas $\R
a b$, $\R b c$, and $\R a c$.  If these atoms have been assigned the
negative polarity, then $\Xi_3$ is the initial rule, and $A$ is $\R a
c$.  Also, $\Xi_1$ and $\Xi_2$ must end with the Release rule.  As a
result, the inference rule constructed here is the following
\emph{backward-chaining} rule:
\[
\infer{\Gamma\vdash \R a c}{\Gamma\vdash \R a b \qquad \Gamma\vdash \R b c}
\]
On the other hand, if these atoms have been assigned the positive
polarity then $\Xi_3$ must end in the Release rule, and $\Xi_1$ and
$\Xi_2$ must be the initial rule, which implies that $\Gamma$ can be
written as $\Gamma',\R a b, \R b c$.  As a result, the inference rule
constructed here is the following \emph{forward-chaining} rule:
\[
\infer{\Gamma',~\R a b,~ \R b c\vdash A}{\Gamma',~\R a b,~\R b c,~\R a c\vdash  A}
\]
The fact that these two choices of polarity for atoms yield these two
styles of inference rules was first published in the papers by
Chaudhuri~\citeyear{chaudhuri06phd} and Chaudhuri et
al.~\citeyear{chaudhuri08jar}.

It is also possible for some atomic formulas to have positive polarity
and some to have negative polarity.  For example, if the atoms $\R a b$
and $\R a c$ have negative polarity and $\R b c$ has positive polarity
then the inference rule built (from the focused derivation above) is
\[
  \infer{\Gamma', \R b c\vdash \R a c}{\Gamma', \R b c \vdash \R a b}
\]
The $\lambda$RCC proof system \cite{jagadeesan05fsttcs} allows for
mixing both forward chaining and backward chaining in a superset of
the hereditary Harrop fragment of intuitionistic logic.  In that proof
system, forward chaining is used to encode constraint propagation as
found in concurrent constraint programming, and backward chaining is
used to encode goal-directed search as found in \lP.  While the
$\lambda$RCC proof system is not a focusing system explicitly, Liang
and the author \citeyear{liang09tcs} showed that it can be accounted
for using \LJF by polarizing the atomic formulas denoting constraints
positive and polarizing the remaining atomic formulas negative.
Chaudhuri \citeyear{chaudhuri10lpar} also used flexible polarity
assignments to model magic set transformations.

Choosing between forward chaining and backward chaining can result in
very different-sized proofs.
Consider, for example, the following specification of the Fibonacci
series as the set $\Pscr$ of three Horn clauses.
\[ \fib{0\;0},\quad\fib{1\;1},\quad\forall n\forall f\forall f'
   [\fib{n\;f}\supset \fib{(n+1)\;f'}\supset \fib{(n+2)\;(f+f')}]
\]
If $f_n$ denotes the $n^{th}$ Fibonacci number then it is easy to
prove that $\fib{n\;m}$ is provable if and only if $m=f_n$ (assuming a
suitable implementation of natural number arithmetic).  The impact of
polarity assignment is on the structure of proofs.  In particular, if
all atomic formulas are made negative, then there exists only
one focused proof of $\fib{n\;f_n}$: this one uses backward chaining,
and its size is exponential in $n$.  On the other hand, if all atomic
formulas are made positive, then there is an infinite number
of focused proofs, all of which use forward chaining: the smallest such
proof has size linear in $n$.

Consider now the paper by Kowalski~\citeyear{kowalski79cacm}
where he proposed the equation
\[\hbox{\em Algorithm}=\hbox{\em Logic}+\hbox{\em Control}.\]
One component for controlling proof search with Horn clauses was
identified in that paper as ``Direction (e.g., top-down or
bottom-up)''.  In the early literature on logic programming, the
connection between top-down and bottom-up search in Horn clauses and
resolution was known to be related to hyper-resolution (for bottom-up)
and SLD-resolution (for top-down).  (See \cite{warren18tplp} for a
description of how these two forms of search have been integrated into
a Prolog system using a tabling mechanism.)
As the discussion in this section
makes clear, this particular component of control now has a rather
elegant proof-theoretic explanation: within a focused proof system,
choose negative polarization for atoms to specify top-down (backward
chaining) or choose positive polarization for atoms to specify
bottom-up (forward chaining).  Choosing a mixture of positive and
negative polarity for atoms yields a mixture of these two
search strategies.

Other aspects of control (of which there are many) are not captured by
focusing classical, intuitionistic, and linear logics.  For example,
the left-to-right ordering of conjunctive goals is not captured by
focusing alone.  For that, there have been some results
surrounding non-commutative logic
\cite{lambek58,retore97tlca,abrusci99apal,polakow99tlca,guglielmi07tocl} and
associated logic programming languages \cite{ruet97csl,polakow01phd}.
Still another aspect of control in logic programming is to allow certain
special goals to be treated as \emph{constraints} that can be delayed
and solved by external solvers \cite{jaffar87popl}.  This approach to
constraints has been effectively implemented in numerous Prolog
systems, such as SWI-Prolog \cite{wielemaker12tplp} and in the Elpi
implementation of $\lambda$Prolog \cite{coen19mscs}.

\section{Advantages for connecting logic programming to proof theory}
\label{sec:3-10}

Using proof theory as a framework for describing and studying logic
programming has at least the following benefits.

\begin{enumerate}
\item This framework has allowed researchers to extend the role of
  logic in logic programming beyond first-order Horn clauses to
  include much richer logics involving higher-order quantification,
  intuitionistic logic, and linear logic.

\item This framework also makes it possible to see the simpler logic
  programs as part of a richer logic (in the survey here, that logic is
  linear logic).

\item Proof theory has also made it possible to vividly 
  compare the nature of functional programming (as proof
  normalization) and logic programming (as proof search).

\item Given the often close relationship between type theory and the
  proof theory of intuitionistic logic, there has been a strong flow
  of design principles and implementation techniques from logic
  programming to type systems: examples of such a flow can be found in
  \cite{pfenning88lfp,elliott89rta,pfenning89lics,felty90cade,pfenning91lf,cervesato96lics}.

\item A satisfactory proof-theoretic treatment of Clark's program
  completion \cite{clark78} was developed in the early 1990s using
  inference rules that worked directly with equality and fixed
  points~\cite{girard92mail,schroeder-heister93lics,mcdowell00tcs}.
  Those innovations allow sequent calculus proof systems to capture
  not only negation-as-finite-failure but also a range of model
  checking problems~\cite{heath19jar}.

\end{enumerate}

Proving that cut elimination holds for a given sequent calculus proof
system is probably the most important meta-theoretical result 
for such a proof system.  The cut-elimination theorem usually implies
the consistency of the logical system described by the proof system,
and it is usually the starting point for describing proof search
strategies.  It has also been used to help in reasoning about logic
programs as well.  For example, \emph{collection analysis} of Horn
clause logic programs can be done statically using linear logic and
cut-elimination \cite{miller08andrews}.  The Abella theorem prover
\cite{baelde14jfr} encodes a two-level logic approach to reasoning
about computation \cite{gacek12jar}.  One of these logic levels is for
the logic programs used to specify computation; the second logic level
captures the first level's metatheory using induction and coinduction.
Two of Abella's tactics are based on the cut-elimination theorem for
the logic specification level.  In many meta-theoretic proofs, these
cut-elimination-based tactics immediately provide proofs of key
substitution lemmas, i.e., lemmas stating that if a certain predicate
holds for a term, it also holds for all instances of that
term~\cite{gacek12jar}.

Another advantage of basing a programming language within proof theory
is that complexity results regarding proof theory can be immediately
applied to logic programs.  In particular, since it is known that any
Turing computable function can be computed using first-order Horn
clauses \cite{tarnlund77}, the first-order fragments of all the logic
programming languages we have seen are undecidable, since they all
include provability in Horn clauses.  When we restrict to
propositional logics, we have the following results: satisfiability
and provability in propositional Horn clauses is linear time
\cite{dowling84jlp}, provability of propositional hereditary Harrop
formulas is polynomial-space complete \cite{statman79tcsa}, and
propositional linear logic is undecidable, even when there are no
propositional variables \cite{lincoln95all}.

In Section~\ref{sec:problems}, we listed several shortcomings of Horn
clause logic programming languages, such as Prolog.  Some of these
shortcomings are addressed, to some degree, by linking logic
programming more closely to proof theory.  Probably the most
significant improvement to the logic programming paradigm is the
inclusion of programming level \emph{abstractions}: as we have seen, the sequent
calculus supports higher-order programming (Section~\ref{sec:ho}),
modular program construction (Section~\ref{sec:modules}), abstract
syntax for data containing bindings (Section~\ref{ssec:ltree}), and
abstract data types (Section~\ref{ssec:adt}).  Making a connection to
linear logic also allows for certain forms of \texttt{assert} and
\texttt{retract} to be provided~(Section~\ref{sec:llp}).  The use of
the proof-theoretic notion of polarization and focused proof has also
provided descriptions of both bottom-up and top-down proof search as
well as combinations of these two (Section~\ref{ssec:fcbc}).

\section{Prospects for logic programming}
\label{sec:prospects}

Logic programs have often been and continue to be deployed to build
various kinds of database systems, interpreters of other languages,
parsers, and type inference engines: for such examples, see the
popular texts \cite{maier88book,okeefe90book,clocksin94}.  Given the
prominence of proof theory in this paper, the following comments on
the prospects for logic programming are limited to those tasks that
demand effective implementation of trustworthy logical deduction.

Traditionally, Prolog has not made a strong commitment to logical
correctness given the large number of non-logical primitives in it,
ranging from assert/retract, to univ, the cut control operator (!),
negation-as-failure, and the absence of the occur-check.  Fortunately,
more recent logic programming systems have put much more focus on
implementing sound logical reasoning.  Systems such as Teyjus
\cite{nadathur99cade,nadathur05tplp}, Elpi \cite{dunchev15lpar},
miniKanren \cite{fiedman18book}, and Makam \cite{stampoulis18icfp},
have made logical soundness a goal, at least for core aspects of their
implementations.

There are many places in the analysis of software, logic, and proof
where logic programming can be applied but where soundness is
critical: we expand on several such topics in the rest of this
section. 

\subsection{Theorem proving}
\label{ssec:tp}

Several of today's interactive theorem provers make use of LCF
\emph{tactics and tacticals}
\cite{milner79mfcs,gordon79,gordon00milner}, which are themselves
generally implemented using higher-order functional programs.  With
the advent of higher-order logic programming languages, such as \lP,
the argument has been made that logic programming would make for a
more flexible and natural setting to implement such tactics and
tacticals, especially since the applications of tactics can fail and
require backtracking \cite{felty89phd,felty93jar}.  More recently, the
Elpi implementation of \lP has been integrated into Coq as a plugin
\cite{tassi18coqpl} and used to help automate aspects of the Coq
prover \cite{tassi19itp}.

Early papers, such as those by Stickel~\citeyear{stickel88jar} and Wos
and McCune~\citeyear{wos91jlp}, point out the rich connections between
logic programming and automated deduction and the cross-fertilizations
of implementation techniques between those two domains.  Some years
later, systems such as leanTAP \cite{beckert95jar,lisitsa03tr} and
leanCop \cite{hodas01ijcar,otten03jsc} were built around the notion of
\emph{lean deduction} in which small Prolog programs were capable of
capturing sound and complete theorem provers for first-order logic.

\subsection{Proof checking}
\label{ssec:checking}

The logic programming paradigm is a natural candidate for performing
proof checking \cite{miller17fac} for several reasons.  First, there
are many kinds of proof certificates in use these days.  In almost all
cases, those certificates do not contain all the
required details to formally check a proof.  Instead, many details are left
implicit, and so the proof checker will, in general, need to perform
some forms of proof reconstruction.  Here, standard logic programming
technology---unification and backtracking search---can be
employed.  For example, a certificate might not contain the actual
substitution terms needed to instantiate a quantifier.  Logic variables
and unification can infer such substitution terms.
Similarly, claiming that a goal formula is already present in the
context requires an index into the context as the witness of that
claim.  Backtracking search can also be used to find such a
witness.  Second, quantificational logic formulas and their proofs often contain
variable bindings to capture both quantifiers and eigenvariables.
There are several logical frameworks, in particular, higher-order
hereditary Harrop formulas (by virtue of being based on Church's STT
\citeyear{church40}) and the LF logical framework \cite{harper93jacm},
that provide a purely logic-based representation of such binding
structures.  Implementations of such frameworks---\lP and Twelf
\cite{pfenning99cade}---treat binding structures via both
unification and backtracking search.

Early use of logic programming to implement proof checkers was
explored within the Proof Carrying Code project
\cite{necula97popl,appel99iclp}.  In that context, logic programming
allowed for compact, flexible, and easy to understand proof checkers.
Logic programming was used as the core motivation of the
\emph{foundational proof certificate} (FPC) project
\cite{chihani17jar}.  In that project, a proof certificate can be seen as
a data structure that incorporates control information for a
simplistic sequent calculus theorem prover.  The FPC framework can be
used as both a kernel itself (assuming that one is willing to admit a
logic programming implementation into the trusted base) or as part of
a toolchain that allows for the flexible manipulation of proof
certificates.  In the latter setting, proof checking can be organized
to transform a proof certificate with some details missing into a
fully detailed proof structure that could be given to existing and
trusted kernels, such as is found in Coq \cite{blanco17cade}.

\subsection{Software systems}
\label{ssec:ss}

If we consider programming as merely the activity of\/ ``writing and
shipping code with the hope that it does not do much harm''---which
characterizes much about programming to date---then it seems unlikely
that logic programming languages will impact the building of software
systems.  However, it seems clear that we should broaden the
discipline of programming to include many other activities that can
improve the quality and correctness of programming.  Such activities
can include automated testing, various kinds of static analyses,
program transformation and refinement, and proving partial or full
functional correctness of code.  Once we add all of these activities
to the programming discipline, then logic programming can play a
sizable role since it has important uses in all of these additional activities.

Logic programming had early successful uses in the specification of
the operational semantics of programming language using either
\emph{structural operational semantics}
\cite{plotkin81,plotkin04jlapb} or \emph{natural
semantics}~\cite{despeyroux86lics,kahn87stacs}.  For example, the
Typol subsystem \cite{despeyroux84,clement85tr,despeyroux88tr} of the
Mentor~\cite{donzeau-gouge84ipe} and Centaur~\cite{borras88} systems
compiled both dynamic and static semantic definitions of various
programming languages into Prolog in order to generate parsers, type
checkers, compilers, interpreters, and debuggers.

Many of the early and most convincing logic programming applications
in higher-order, intuitionistic logic involved the mechanization of
the meta-theory of functional programming
\cite{hannan90phd,michaylov91,hannan92mscs,hannan93jfp}.  Verifiable
compilers have been described and implemented in Elf
\cite{hannan92lics} and in \lP
\cite{whalen05phd,wang16phd,wang16esop}. 

\subsection{Reasoning directly with logic programs}
\label{ssec:reasoning}

Given that the logic programming specifications used to encode
programming language semantics and inference rules are concise and
based on logic itself, there should be rich ways to reason on such
specifications directly.  In the context of the Typol system, such
reasoning could be done by treating provable atomic goals as belonging
to an inductive data type~\cite{despeyroux86lics}.  A more
sophisticated approach to reasoning directly on logic programming has
been developed within the Abella theorem prover \cite{baelde14jfr}.
That prover includes such innovations as the
$\nabla$-quantifier~\cite{miller05tocl,gacek11ic} and the two-level
logic approach to reasoning \cite{gacek12jar}.  As we mentioned above,
the cut-elimination result for the object-logic (the logic programming
specification) is turned into a proof technique in Abella for
reasoning about such logic specifications.

\subsection{A defense of declarative techniques}
\label{ssec:defense}

One advantage of having a proof theory for logic programming is that
it sometimes makes it possible to write compact, high-level
specifications for which correctness is easy to establish.  At the
same time, techniques such as partial evaluation~\cite{lloyd91jlp},
program transformation~\cite{pettorossi94jlp}, and various forms of
static analysis can often be applied directly to specifications
written using logical expressions.  As a result, rich manipulations of
specifications are possible.

As an example of how such manipulations can be applied to logic
specifications in a rich programming language, consider the following
example, taken from \cite{hannan92mscs}.  The specification of
call-by-name evaluation of the untyped $\lambda$-terms can be given as
a binary relation using two higher-order Horn clauses and two
constructors (encoding the untyped $\lambda$-terms).  Given its
simplicity, the correctness of that specification is easy to
establish.  Since that specification is written as logical formulas, a
sequence of transformations can be applied to that specification until
it results in the specification of an \emph{abstract machine} in which
an argument stack and De Bruijn numerals~\cite{debruijn72} are used
encode $\lambda$-terms.  This latter specification can be written
using only first-order (binary) Horn clauses.  Given the correctness
of the initial specification and correctness of the transformations
used, the correctness of the derived low-level specification---which
has an effective implementation in Prolog---easily follows.

Similar examples can be found in \cite{cervesato98jicslp}, where
aspects of the Warren abstract machine were developed by the direct
manipulation of higher-order logic specifications and in
\cite{pientka02iclp}, where proof theory techniques helped to design a
strategy for tabled evaluation of (higher-order) logic programs.

\subsection{Further advances in proof theory}
\label{ssec:advances}

The relationship between logic programming and proof theory is not
just in one direction.  The author has documented in \cite{miller21pt}
several influences of logic programming research on structural proof
theory.  One computational feature that is often desired in the logic
programming world is \emph{saturation}: that is, one would like to
know that forward chaining from a given set of clauses will not yield
new atomic facts being derived.  Saturation was a key component of the
Gamma multiset rewriting programming language \cite{banatre96cp} and
the work on \emph{logical
algorithms}~\cite{ganzinger01ijcar,ganzinger02iclp,simmons08icalp}.
Currently, structural proof theory does not appear to have any
techniques that can account for saturation.

Answer set programming (ASP), as described in papers by Brewka et
al.~\citeyear{brewka11cacm} and Lifschitz~\citeyear{lifschitz08aaai},
is a form of declarative programming that describes computation as the
construction of stable models~\cite{gelfond88iclp}.  The operational
semantics behind such search resembles Datalog's bottom-up inference
along with saturation and the negation-as-failure approach to
negation.  While proof structures exist in this domain (see, for
example, \cite{marek93book} and \cite{lifschitz96csli}), those
structures are seldom related to the proof structures found in
structural proof theory (which has been our focus here).  Some of the
proof-theoretic topics described in Sections~\ref{ssec:fcbc}
and~\ref{sec:3-10} might also relate proof structures to ASP.
Schubert and Urzyczyn \citeyear{schubert18tplp} have considered
initial steps in that direction.

Finally, developing model-theoretic semantics for these rich
proof-theory-inspired languages is interesting to considered.  Lipton
and Nieva~\citeyear{lipton18tcs} have shown how to extend the Kripke
$\lambda$-models of Mitchell and Moggi~\citeyear{mitchell91apal} to
treat an extension to higher-order hereditary Harrop with constraints.
A Kripke-style model for Lolli was given in \cite{hodas94ic}.
While Girard~\citeyear{girard87tcs} considered various forms of
model theory semantics for linear logic, his models have been hard to
apply directly to logic programming: an exception is the paper by
Fages et al.~\citeyear{fages01ic}.

\section{Conclusion}
\label{sec:conclusion}

Structural proof theory has played an essential role in understanding
the nature and structure of logic programming languages.  This role has
been significant when one wants to have more expressive,
dynamic, and modern versions of Prolog.  The proof theory of
first-order and higher-order versions of intuitionistic and linear
logics have provided designs for logic programming languages that
support higher-order and modular programming, abstract data-types, and
state.  Additionally, the theory of focused proofs provides a
satisfying description of how to specify forward chaining and backward
chaining during proof search.

In 1991, Peter Schroeder-Heister \citeyear{schroeder-heister91jsl} and
the author \citeyear{miller91alp} (independently) wrote opinion pieces
in which they proposed that the sequent calculus was an appropriate
framework for exploring the semantics of logic in philosophical and
computational settings.  The goal of those papers was to ensure that
the term ``semantics'' was not just understood in terms of model
theory and denotational semantics.  This survey outlines the successes
and methods that have arisen from using proof theory based on the
sequent calculus as a semantic framework for logic programming.

\medskip
\noindent{\bf Acknowledgments:} I thank Miroslaw Truszczynski and
several anonymous reviewers for their comments on an earlier version
of this paper.

\bibliographystyle{acmtrans}

\label{lastpage}
\end{document}